\documentclass[
twocolumn, 
superscriptaddress,
amsmath, amssymb, aps,
prb,
longbibliography
]{revtex4-2}

\usepackage[dvipdfmx]{graphicx}
\usepackage{dcolumn}
\usepackage{bm}
\usepackage{hyperref}
\usepackage{xcolor}

\usepackage{txfonts}
\usepackage{color} 
\usepackage{listings}

\hypersetup{
  colorlinks=true,
  linkcolor=[rgb]{0.60,0.00,0.00},
  citecolor=[rgb]{0.00,0.00,0.60},
  urlcolor=[rgb]{0.00,0.00,0.60},
  setpagesize=false
}

\begin{document}
    \title{Efficient variational quantum circuit structure for correlated topological phases}
    
    \author{Rong-Yang Sun}
    \affiliation{Computational Materials Science Research Team, RIKEN Center for Computational Science (R-CCS), Kobe, Hyogo, 650-0047, Japan}
    \affiliation{Quantum Computational Science Research Team, RIKEN Center for Quantum Computing (RQC), Wako, Saitama, 351-0198, Japan}
    \author{Tomonori Shirakawa}
    \affiliation{Computational Materials Science Research Team, RIKEN Center for Computational Science (R-CCS), Kobe, Hyogo, 650-0047, Japan}
    \affiliation{Quantum Computational Science Research Team, RIKEN Center for Quantum Computing (RQC), Wako, Saitama, 351-0198, Japan}
    \author{Seiji Yunoki}
    \affiliation{Computational Materials Science Research Team, RIKEN Center for Computational Science (R-CCS), Kobe, Hyogo, 650-0047, Japan}
    \affiliation{Quantum Computational Science Research Team, RIKEN Center for Quantum Computing (RQC), Wako, Saitama, 351-0198, Japan}
    \affiliation{Computational Quantum Matter Research Team, RIKEN Center for Emergent Matter Science (CEMS), Wako, Saitama 351-0198, Japan}
    \affiliation{Computational Condensed Matter Physics Laboratory, RIKEN Cluster for Pioneering Research (CPR), Saitama 351-0198, Japan}

    \begin{abstract}
        We propose an efficient circuit structure of variational quantum circuit \textit{Ans\"{a}tze} used for the variational quantum 
        eigensolver (VQE) 
        algorithm in calculating gapped topological phases on the currently feasible noisy intermediate-scale quantum computers. 
        An efficient circuit \textit{Ansatz} should include two layers: the initialization layer and the variational layer. In the initialization layer, 
        a fixed depth circuit state with a compatible entanglement structure to the target topological phase is constructed. 
        The circuit state is further adjusted subsequently to capture the details of the local correlations, which is dictated 
        with the Hamiltonian, in the parametrized variational layer. 
        Based on this strategy, we design a circuit \textit{Ansatz} to investigate the symmetry-protected 
        topological Haldane phase in a \textit{non-exactly} solvable alternating spin-$1/2$ Heisenberg chain by VQE calculations. 
        Main characterizations of the Haldane phase, including the long-ranged string order, the four-fold nearly degenerate 
        ground states 
        associated with four different localized edge mode patterns for the system with open boundaries, and the two-fold degeneracy of the 
        entanglement 
        spectrum, are all observed for the optimized shallow circuit state with only one depth variational layer
        both in numerical simulations and on real quantum computers.
        We further demonstrate that the computational capacity (i.e., expressibility)
        of this quantum circuit \textit{Ansatz}
        is determined not by the system size but only 
        by the intrinsic correlation length of the system, thus implying that the scalable VQE calculation is possible. 
    \end{abstract}

    \date{\today}

    \maketitle

    \section{Introduction\label{sec:intro}}
    
    Exploring exotic phenomena that emerge in quantum many-body systems is one of the prominent issues in modern condensed 
    matter physics. However, because of its exponentially increased complexity, 
    one still lacks 
    an efficient theoretical tool to investigate these systems in general. Quantum simulation~\cite{georgescu2014quantum}, 
    which has been proposed for a long time~\cite{feynman1982simulating}, provide a promising solution by realizing them in controllable 
    synthetic quantum systems and exploring their properties. Nevertheless, realizing such a well-controlled artificial quantum device is 
    also an extremely challenging task~\cite{altman2021quantum}.

    Recently, quantum simulation has attracted increasing interest since the successful realization of the analog 
    quantum simulators in cold atom systems~\cite{gross2017quantum,schafer2020tools,takahashi2022quantum} and the digital 
    quantum simulators using programmable superconducting transomons~\cite{arute2019quantum} and trapped 
    ions~\cite{pogorelov2021compact}. Specially, the latter ones, i.e., circuit-based quantum computers, have received particular 
    attention because of their flexibility to universal quantum computing, although a generic quantum circuit with any depth cannot be 
    evolved with high fidelity on the current implementation, hence regarded as the noisy intermediate-scale quantum (NISQ) 
    device~\cite{preskill2018quantum}. On the other hand, the algorithms suitable for the NISQ device, which usually involves 
    shallow circuits, are currently under investigation~\cite{bharti2022noisy}.

    Many interesting phases of matter have been realized on the NISQ device, such as a symmetry-protected topological (SPT) 
    phase~\cite{azses2020identification,smith2022crossing} and a topological quantum spin liquid~\cite{satzinger2021realizing}, by 
    \textit{precisely} constructing these states in shallow quantum circuits. These exact circuit states usually correspond to some 
    exquisitely exactly solvable Hamiltonians, such as the cluster model~\cite{verresen2017one}
    (realized in Refs.~\cite{azses2020identification,smith2022crossing}) and the Kitaev toric code model~\cite{kitaev2003fault} 
    (realized in Ref.~\cite{satzinger2021realizing}). However, it is usually impossible to obtain the exact solution  
    of a more generic and realistic many-body Hamiltonian.

    Generally, a quantum many-body state, considered as the ground state of some complex Hamiltonian, can be prepared 
    on the circuit-based quantum computer by performing digitized adiabatic quantum computing (AQC) 
    process~\cite{Barends2015,Barends2016,mbeng2019optimal} (for a review, see Ref.~\cite{albash2018adiabatic}), 
    which is also known as quantum annealing (QA)~\cite{kadowaki1998quantum} or adiabatic state preparation 
    (ASP)~\cite{alan2005simulated}. However, the AQC usually requires deep circuits, thus challenging for the present NISQ device. 
    Following the same spirit of the quantum approximate optimization algorithm~\cite{Farhi2014}, 
    the quantum adiabatic path itself can be optimized variationally to obtain the desired state with a relatively shallow 
    circuit~\cite{Ho2019,Mbeng2019a,Mbeng2019b,Wauters2020,Shirakawa2021}, and various auxiliary techniques such as circuit 
    recompiling~\cite{sun2021quantum,tan2021realizing} and counterdiabatic 
    driving~\cite{Campo2013,Sels2017,Claeys2019,Xie2022} have also been proposed.

    Apart from preparing a certain quantum state, we can directly evaluate the ground state of the desired quantum many-body 
    Hamiltonian $H$ by solving the related eigenvalue problem on the NISQ device. One of the most extensively studied 
    algorithms for this purpose 
    is the variational quantum eigensolver (VQE)~\cite{peruzzo2014variational,yung2014transistor, cerezo2021variational}. 
    In the VQE scheme, a variational circuit \textit{Ansatz} $|\Psi(\boldsymbol{\theta})\rangle$ parametrized with the 
    variational parameters $\boldsymbol{\theta}=(\theta_1,\theta_2,\theta_3,\dots)$ is evolved on a quantum computer to 
    estimate the associated variational energy 
    $E(\boldsymbol{\theta}) = \langle \Psi(\boldsymbol{\theta})|H|\Psi(\boldsymbol{\theta})\rangle$ as a cost function  
    (for a considerable system, it is almost unattainable on a classical computer), and the variational parameters 
    $\boldsymbol{\theta}$ are updated on a classical computer. This quantum-classical loop is repeated until 
    the variational parameters $\boldsymbol{\theta}$ converge so as to minimize the variational energy $E(\boldsymbol{\theta})$. 
    The resulting $|\Psi(\boldsymbol{\theta}^*)\rangle$ with the optimized variational parameters $\boldsymbol{\theta}^*$ 
    can be considered as a good approximation for the ground state of $H$.

    In principle, the VQE scheme asserts to solve the ground state (as well as excited states) 
    for any quantum many-body Hamiltonian. Nevertheless, the practical 
    applications are still questioned. One of the main issues is how to design the variational circuit \textit{Ansatz} that can be evolved 
    on the present NISQ device with high fidelity. High fidelity implies that the circuit should be designed with a shallow depth. 
    However, the circuit depth usually diverges with the problem size in the circuit \textit{Ansatz} currently applied widely, 
    such as quantum alternating operator \textit{Ansatz}~\cite{hadfield2019from}, which is also inspired by the Trotterized 
    quantum adiabatic transformation.

    Here, we propose an efficient variational quantum circuit structure for the VQE calculations of the ground state 
    in a gapped topological phase.  
    The efficient variational quantum circuit consists of an initialization layer where the basic entanglement structure 
    of the targeted topological phase is built, 
    followed by a variational layer which is adapted to fit the model-parameter-specific fine correlation structure of the ground state. 
    We employ this \textit{Ansatz} strategy to explore the SPT Haldane phase in a realistic, not exactly solvable model, 
    i.e., an alternating Heisenberg chain (AHC). 
    We demonstrate that a nearly exact ground state can be obtained by a very shallow parametrized circuit \textit{Ansatz} in a finite range 
    of the model parameter space. Furthermore, the four-fold nearly degenerate 
    ground states associated with different edge mode patterns are 
    obtained in the circuit states by specifying the initializations at the edges of the system. 
    By implementing these circuit states on IBM quantum computers, we reveal that all the characteristic features 
    of the SPT Haldane phase, 
    including the string order parameter, edge modes, and two-fold degenerate entanglement spectrum, can be clearly observed 
    on the real devices. 
    Moreover, we numerically demonstrate 
    that (i) a deeper variational layer is necessary 
    if the entanglement structure 
    is not correctly set up in the initialization layer 
    and (ii) the computational capacity of an appropriately constructed circuit \textit{Ansatz} is determined by the correlation length 
    of the system, not by the system size. 
    Finally, we discuss that this circuit \textit{Ansatz} has the potential to achieve the quantum advantage on the current 
    NISQ device.

    The rest of the paper is organized as follows. We first briefly summarize the relation between quantum entanglement 
    and a topological phase, and propose an efficient variational circuit structure in Sec.~\ref{sec:eff_circ_struc}. 
    Next, we introduce the AHC model, which is a realization of the SPT Haldane phase in a spin-$1/2$ system, 
    and summarize the main characterizations of the SPT Haldane phase in Sec.~\ref{sec:ahc_model}. 
    In Sec.~\ref{sec:vqe_setup}, we develop a specific variational circuit \textit{Ansatz} for the VQE calculations of the AHC model   
    and examine its symmetry properties. 
    We also explain briefly
    the VQE scheme employed in this study, including the optimization method and the parameter setup. 
    In Sec.~\ref{sec:haldane_phase}, using both numerical simulations and real quantum devices, 
    we characterize 
    the features of 
    the VQE optimized circuit states. 
    In Sec.~\ref{sec:expressibility}, we discuss expressibility of the circuit \textit{Ansatz}. 
    Finally, we conclude this paper by summarizing our results, and discuss several further extensions 
    and the relevance to the quantum advantage.
    Degeneracy of entanglement spectrum of the parametrized circuit states introduced in this study is further examined 
    in Appendix~\ref{app:entanglement}.
    The circuit \textit{Ansatz} depth dependence of the string order in the VQE simulations of the SPT Haldane phase 
    is discussed in Appendix~\ref{app:str_order_D}, and
    the comparison with a generic circuit \textit{Ansatz} without elaborate structures is also provided in 
    Appendix~\ref{app:vs_so4_brick_wall}.

    \section{Efficient variational circuit structure\label{sec:eff_circ_struc}}
    
    \subsection{Quantum entanglement and topological phase}
    
    Exotic quantum phases, which cannot be described by the Landau's symmetry-breaking paradigm and are not characterized 
    by local order parameters, are usually considered to hold topological orders~\cite{wen1990topological,gu2009tensor}. 
    One of the primary characteristics of a topological phase is the nontrivial quantum entanglement structure, which 
    distinguishes it from a trivial direct-product state~\cite{zeng2019quantum}. The nontrivial entanglement structure implies that 
    it cannot be easily disentangled to a direct-product state by local unitary transformations. 
    Therefore, the intrinsic (symmetry-protected) topological phase can be defined, in an operational way, 
    as a state that cannot be connected to a direct-product state by any (certain symmetric) \textit{finite-depth} quantum circuit 
    composed of local gates in the thermodynamic limit. On the other hand, two states can be connected 
    by a quantum circuit with finite depth in the thermodynamic limit when they are in the same topological phase~\cite{chen2010local}.

    Generating a topological state in a quantum computer can be considered as the inverse procedure of the above 
    operational definition. On the digital circuit-based quantum computer, we start from a direct-product initial state, e.g.,  
    $|0,0,\cdots,0\rangle$, and apply a quantum circuit to evolve the initial state to a topologically nontrivial final state.
    Therefore, it is generally expected that the depth of a quantum circuit increases with the system size in order to prepare an intrinsic 
    topological phase~\cite{satzinger2021realizing} or an SPT phase (considering a symmetry preserved 
    quantum circuit)~\cite{tan2021realizing}.

    \begin{figure}[ht]
      \centering
      \includegraphics[width=0.98\linewidth]{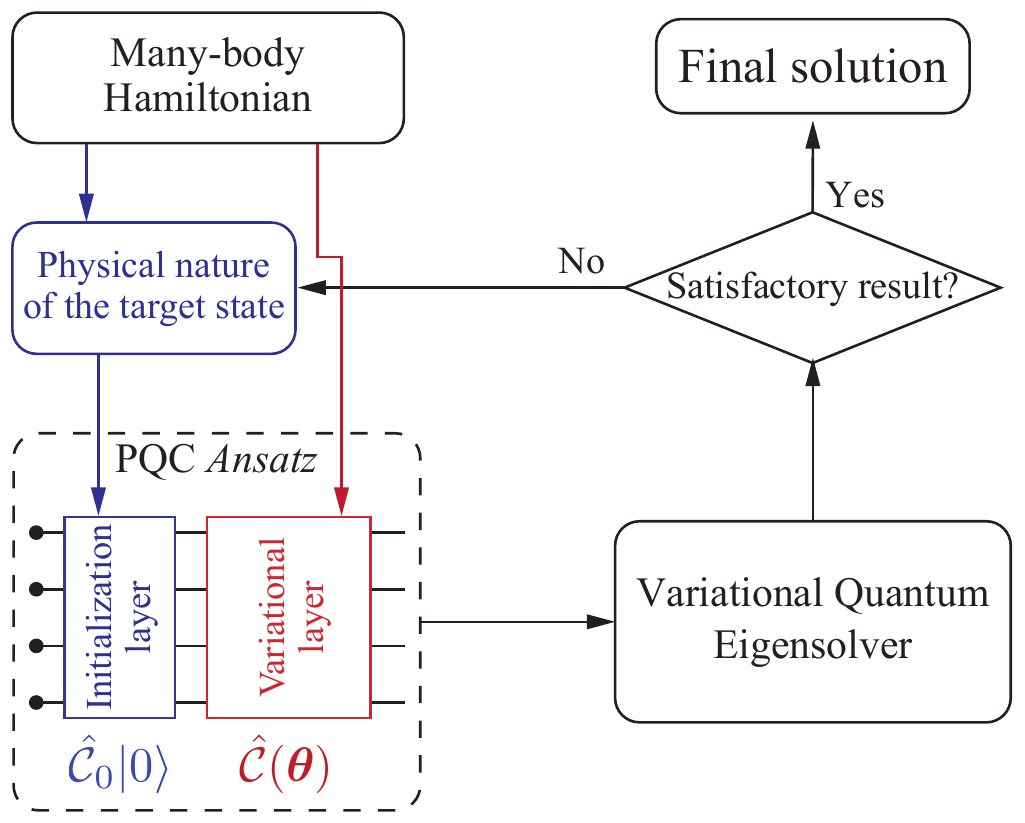}
      \caption{\label{fig:flow_chart}
        Flow chart of the calculation scheme proposed in this study. The lower left block demonstrates 
        the parametrized quantum circuit (PQC) \textit{Ansatz} 
        comprising the initialization and variational layers inspired by the topological nature of the target state and the local details 
        of the Hamiltonian, respectively.}
    \end{figure}

    \subsection{Efficient circuit structure for a topological phase}
    
    From another perspective, the operational definition of a topological phase described above provides an idea on 
    how to design an efficient variational circuit \textit{Ansatz} to represent a topological state. As the first step, we construct 
    a quantum circuit (as shallow as possible), 
    which can transform the initial product state to a state sharing the same nontrivial topological characterization,
    i.e., in the same topological phase, as the desired final state. This part of the circuit is referred to as the initialization layer 
    hereafter. 
    For example, inspired by the idea of the renormalization group, the fixed-point state of the topological phase with zero correlation 
    length is a suitable choice to be constructed in the initialization layer, 
    such as the toric code state~\cite{aguado2008entanglement} and the cluster state~\cite{smith2022crossing}.
    However, a quantum circuit state representing the ground state of 
    a more generic Hamiltonian cannot be obtained unless its local correlations are further 
    incorporated. We hence need a second step, in which a parametrized circuit, referred to as the variational layer, is introduced to correct 
    the quantum circuit state generated in the initialization layer. Since the quantum circuit state generated in the initialization layer 
    already has the same entanglement structure as the desired topological phase, we expect that the depth of the variational layer 
    does not increase with the system size. This is indeed the case, as will be demonstrated in Sec.~\ref{sec:expressibility}.

    Therefore, an efficient variational circuit \textit{Ansatz} has the form
    \begin{equation}
        \label{eq:circ_ansa}
        |\Psi(\boldsymbol{\theta})\rangle = \hat{\mathcal{C}}(\boldsymbol{\theta}) \hat{\mathcal{C}}_{0}|0\rangle~,
    \end{equation}
    where $\hat{\mathcal{C}}_{0}$ is the initialization layer and $\hat{\mathcal{C}}(\boldsymbol{\theta})$ represents the variational layer 
    with a set of variational parameters $\boldsymbol{\theta}$. 
    $|0\rangle$ is the initial state where all the quantum registers are set to be zero. 
    While the depth of $\hat{\mathcal{C}}_{0}$ increases with the system size if the target final state is 
    an intrinsic topological phase, it is independent of the system size if 
    the target final state is an SPT phase, assuming that the symmetry protecting its nontrivial topology of the SPT phase 
    is broken in the initialization layer. 
    On the other hand, the depth of $\hat{\mathcal{C}}(\boldsymbol{\theta})$ is always finite and does not scale with the system size. 
    Indeed, the number $D$ 
    of layers in the circuit $\hat{\mathcal{C}}(\boldsymbol{\theta})$ is determined by the local correlation length of the system.
    Figure~\ref{fig:flow_chart} illustrates a flow chart of the calculation scheme to utilize this circuit \textit{Ansatz} for 
    simulating 
    topological phases, which will be employed in the following sections.

    The circuit \textit{Ansatz} in Eq.~(\ref{eq:circ_ansa}) can also be regarded as an AQC process when the initialization layer constructs 
    the ground state of a Hamiltonian, for example, the fixed point Hamiltonian, which is adiabatically connected 
    to the final target Hamiltonian by tuning some model parameters. However, we should emphasize that the quantum circuit state 
    constructed in the initialization layer can also be independent of any related microscopic Hamiltonians and can be chosen as, e.g., a 
    topological spin liquid state generated by projective construction~\cite{wen2002quantum}.

    \section{SPT Haldane phase in a spin-1/2 system\label{sec:ahc_model}}
    
    The original Haldane phase in the spin-$1$ Heisenberg chain is hard to be implemented straightforwardly on most 
    of the currently available quantum computing platforms since a single quantum bit is based on a spin-$1/2$ object. 
    Nevertheless, if the Hamiltonian is allowed to break partially the translational symmetry, the Haldane phase can also exist 
    in a spin-$1/2$ chain. One of the examples is the AHC~\cite{hida1992crossover}, which is considered in this paper.

\begin{figure}[t]
        \centering
        \includegraphics[width=\linewidth]{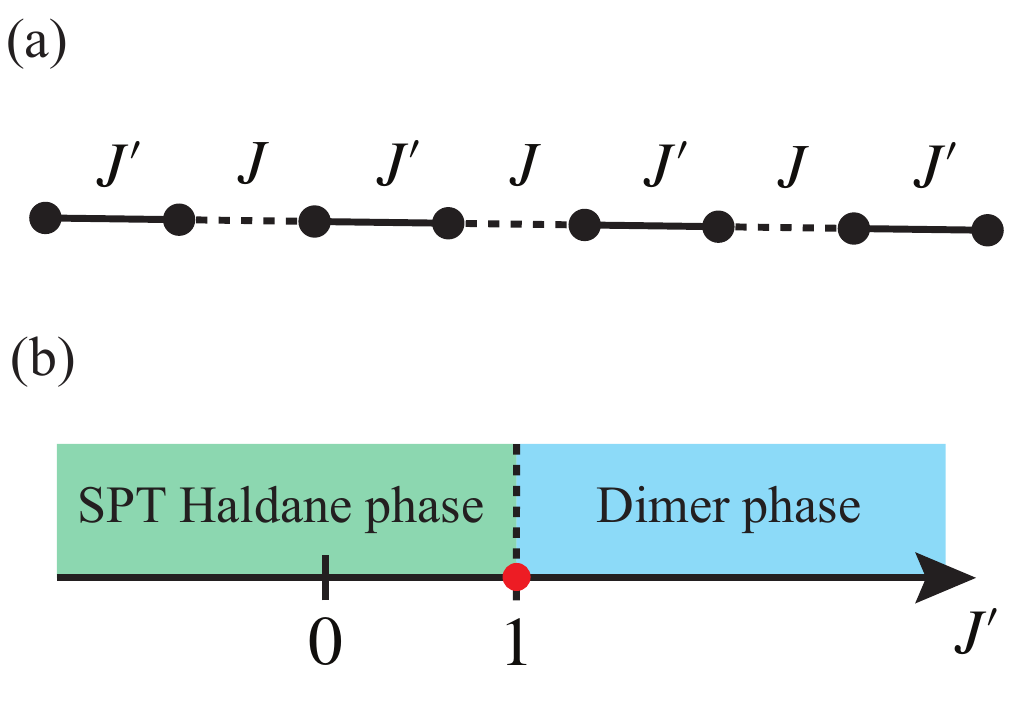}
        \caption{\label{fig:model}
            (a) Schematic figure of the alternating Heisenberg chain (AHC). Here, we assume $L=8$ under open boundary conditions. 
            Lattice sites are indicated by solid circles and the intra-unitcell (inter-unitcell) coupling $J'$ ($J$) is denoted 
            by solid (dashed) lines. 
            (b) Schematic phase diagram for the ground state of the AHC, setting $J=1$ as the energy unit. 
        }
    \end{figure}

    The AHC is described by the following Hamiltonian:
    \begin{equation}
        \label{eq:alth_ham}
        H = \sum_{i = 0}^{L / 2-1} \left( J'\mathbf{S}_{2i} \cdot \mathbf{S}_{2i+1} + J\mathbf{S}_{2i+1} \cdot \mathbf{S}_{2(i+1)}\right),
    \end{equation}
    where $L$ is the number of sites, assumed to be a multiple of 4, and $\mathbf{S}_{k}$ is the spin-$1/2$ operator located at site $k$, 
    enumerated from 0 to $L-1$ with $\mathbf{S}_L=\mathbf{S}_0$ under periodic boundary conditions. 
    Considering that neighboring two spins at sites $2i$ and $2i+1$ form an unitcell labeled by $i$, 
    the number of unitcells is thus even for $L$ being a multiple of 4 
    and $J'$ ($J$) is the intra-unitcell (inter-unitcell) spin coupling [see Fig.~\ref{fig:model}(a)].  
    Without losing generality, 
    we fix $J = 1$ as the energy unit in this paper.

    The ground state phase diagram of the AHC has been well studied~\cite{hida1992crossover} and is shown schematically 
    in Fig.~\ref{fig:model}(b). 
    At the $J' = 1$ point, the Hamiltonian $H$ in Eq.~(\ref{eq:alth_ham}) is restored to the isotropic spin-1/2 antiferromagnetic 
    Heisenberg chain, for which the ground state has a gapless spin excitation and exhibits power-law decay of the spatial correlation 
    function. 
    For $J' > 1$, the ground state is in a spin-singlet dimer phase (i.e., a singlet dimer formed in each unitcell) with a trivial gap. 
    The fixed point of this trivial dimer phase is at $J' = \infty$, where no correlation is built between the neighboring unitcells. 
    For $J' < 1$, the ground state is a gapped SPT phase sharing the same nature as the spin-1 Haldane phase~\cite{hida1992crossover}. 
    The AHC is equivalent to the spin-1 Haldane chain at the limit of $J' = -\infty$, where two spins in the unitcell are grouped to form an 
    effective spin-1 spin with an effective inter-unitcell spin coupling $J/4$. 
    This SPT Haldane phase is protected by either the $Z_{2}\times Z_{2}$ symmetry 
    with respect to $\pi$ rotations referring to a pair of orthogonal axes~\cite{pollmann2012symmetry}, the time-reversal symmetry, 
    or the bond (in the AHC case, the bond connecting two neighboring unitcells) centered inversion 
    symmetry~\cite{gu2009tensor, pollmann2010entanglement}. The fixed point of this SPT phase is at $J' = 0$.

    To characterize the nontrivial topological property of the SPT Haldane phase, similar to the case for the spin-1 model, 
    we can define the corresponding non-local string operator for the AHC:
    \begin{equation}
        \label{eq:str_op}
        \mathcal{O}_{\rm str}(d) = \mathcal{S}^{z}_{k} \left(\prod_{l=k+1}^{k+d-1} \exp(i\pi \mathcal{S}^{z}_{l})\right) \mathcal{S}^{z}_{k+d}~,
    \end{equation}
    where the spin operator $\mathcal{S}^{z}_{k}$ represents the $z$ component of the total spin in the $k$-th unitcell, i.e., 
    $\mathcal{S}^{z}_{k} = S^{z}_{2k} + S^{z}_{2k+1}$. The string order parameter $O_{\rm str}$ can be defined as 
    $O_{\rm str} \equiv \lim_{d\rightarrow \infty} \langle\mathcal{O}_{\rm str}(d)\rangle$ in the thermodynamic limit, where 
    $\langle\cdots\rangle$ implies the expectation value over the ground state. 
    For a finite system, we can evaluate the expectation value of the string operator at the longest reachable distance $d$ 
    in the system, approximating the string order parameter. 
    In addition, for the system under open boundary conditions (OBC), 
    localized edge modes provide a four-fold nearly degenerate ground state, 
    which is also an essential characterization of the Haldane phase. 
    Moreover, this SPT phase is further characterized uniquely by the two-fold degeneracy of its entanglement 
    spectrum~\cite{pollmann2010entanglement,pollmann2012symmetry}. To demonstrate all these three main features 
    of the SPT Haldane phase, we will focus on the finite AHC under OBC [see Fig.~\ref{fig:model}(a)] in the following calculations.

    \section{VQE setup details\label{sec:vqe_setup}}
    
    Following the strategy for the efficient circuit construction described in Sec.~\ref{sec:eff_circ_struc}, here we first design 
    a variational quantum circuit \textit{Ansatz} with eSWAP gates for the VQE calculations of the AHC. 
    We then discuss comprehensively the symmetry of this \textit{Ansatz} to verify that it is appropriate for the calculation of the AHC. 
    Furthermore, we briefly explain the natural gradient descent optimization method with its parameter setup, 
    which will be employed in the VQE calculations given in Sec.~\ref{sec:numerical}.

    \subsection{Variational quantum circuit \textit{Ansatz}} \label{sec:ansatz}
    
    Inspired by the circuit structure in Eq.~(\ref{eq:circ_ansa}), we construct the quantum circuit \textit{Ansatz} for the AHC 
    by separately designing the initialization layer and the variational layer. For the initialization layer, we consider the following 
    six circuit states:
    \begin{eqnarray}
        \hat{\mathcal{C}}^{s}_{0}|0\rangle &=& |s\rangle_{1,2} |s\rangle_{3,4} \cdots |s\rangle_{L-1,0} \equiv |\phi_s\rangle \label{eq:init_lays_s}\\ 
        \hat{\mathcal{C}}^{00}_{0}|0\rangle &=& |0\rangle_0 |s\rangle_{1,2} |s\rangle_{3,4} \cdots |s\rangle_{L-3,L-2} |0\rangle_{L-1} \equiv |\phi_{00}\rangle \label{eq:init_lays_00}\\ 
        \hat{\mathcal{C}}^{01}_{0}|0\rangle &=& |0\rangle_0 |s\rangle_{1,2} |s\rangle_{3,4} \cdots |s\rangle_{L-3,L-2} |1\rangle_{L-1} \equiv |\phi_{01}\rangle \label{eq:init_lays_01}\\ 
        \hat{\mathcal{C}}^{10}_{0}|0\rangle &=& |1\rangle_0 |s\rangle_{1,2} |s\rangle_{3,4} \cdots |s\rangle_{L-3,L-2} |0\rangle_{L-1} \equiv |\phi_{10}\rangle \label{eq:init_lays_10}\\ 
        \hat{\mathcal{C}}^{11}_{0}|0\rangle &=& |1\rangle_0 |s\rangle_{1,2} |s\rangle_{3,4} \cdots |s\rangle_{L-3,L-2} |1\rangle_{L-1} \equiv |\phi_{11}\rangle \label{eq:init_lays_11}\\ 
        \hat{\mathcal{C}}^{d}_{0}|0\rangle &=& |s\rangle_{0,1} |s\rangle_{2,3} \cdots |s\rangle_{L-2,L-1} \equiv |\phi_d\rangle, \label{eq:init_lays_d}
    \end{eqnarray}
    where 
    $|0\rangle_i$ ($|1\rangle_i$) is the local state at qubit $i$, corresponding to site $i$ in the AHC, 
     with $\hat{Z}_i|0\rangle_i=|0\rangle_i$ 
    ($\hat{Z}_i|1\rangle_i=-|1\rangle_i$) and 
    $\hat{Z}_i$ being the Pauli Z gate at qubit $i$, and 
    $|s\rangle_{i,j}$ indicates a singlet state formed by two spin-$1/2$ spins located at sites $i$ and $j$. 
    The first five initialization layers initialize the circuit state to a product of singlets connecting two neighboring unitcells 
    (except for $|s\rangle_{L-1,0}$ being associated with sites $0$ and $L-1$), 
    which are all the ground states of the Hamiltonian in Eq.~(\ref{eq:alth_ham}) with $J' = 0$ 
    (i.e., the fixed point state for the SPT Haldane phase), recalling that 
    the singlet pair formed at sites 0 and $L-1$ across the boundaries in $\hat{C}_0^s|0\rangle$, 
    referred to as $|\phi_s\rangle$, does not affect the energy expectation value under OBC. 
    Notice also that among the states generated in the initialization layers 
    $\hat{C}_0^{00}|0\rangle$, $\hat{C}_0^{01}|0\rangle$, $\hat{C}_0^{10}|0\rangle$, and $\hat{C}_0^{11}|0\rangle$, 
    simply referred as $|\phi_{00}\rangle$, $|\phi_{01}\rangle$, $|\phi_{10}\rangle$, and $|\phi_{11}\rangle$, respectively, 
    only the local states at the boundary sites 0 and $L-1$ 
    are distinct and $|\phi_s\rangle$ can be indeed obtained by a linear combination of $|\phi_{10}\rangle$ 
    and $|\phi_{01}\rangle$, i.e., $|\phi_s\rangle=\frac{1}{\sqrt{2}}(|\phi_{10}\rangle-|\phi_{01}\rangle)$. 
    In contrast, the last one $\hat{C}_0^{d}|0\rangle$ in Eq.~(\ref{eq:init_lays_d}), referred to as $|\phi_d\rangle$, 
    initializes the circuit state to a product of singlets within each unitcell, 
    which is the $J' = \infty$ ground state, i.e., the fixed point state for the trivial dimer phase.

\begin{figure*}[hbtp]
        \centering
        \includegraphics[width=\linewidth]{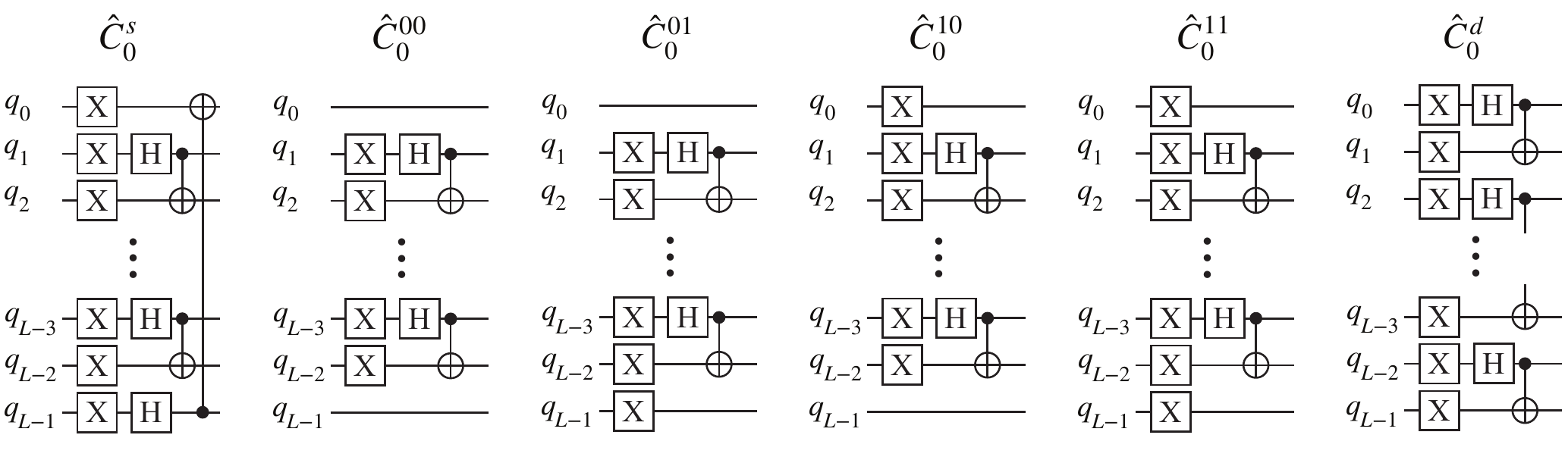}
        \caption{\label{fig:schm_fig}
            Six different initialization layers, $\hat{C}^s_0$, $\hat{C}^{00}_0$, $\hat{C}^{01}_0$, $\hat{C}^{10}_0$, $\hat{C}^{11}_0$, and 
            $\hat{C}^d_0$, introduced in Eqs.~(\ref{eq:init_lays_s})--(\ref{eq:init_lays_d}). 
            Here, all the quantum registers $\{q_0, q_1,\cdots,q_{L-1}\}$ are initially set to be zero to generate the states $|\phi_s\rangle$, 
            $|\phi_{00}\rangle$, $|\phi_{01}\rangle$, $|\phi_{10}\rangle$, $|\phi_{11}\rangle$, and $|\phi_d\rangle$.}
    \end{figure*}

    In detail, the spin singlet $|s\rangle_{i,j}$ formed at sites $i$ and $j$, which appears in 
    Eqs.~(\ref{eq:init_lays_s})--(\ref{eq:init_lays_d}), can be generated by the two-qubit unitary 
    $\hat{U}^{s}_{ij}$, i.e., 
    \begin{equation}
     |s\rangle_{i,j} = \hat{U}^{s}_{ij}|0\rangle_i   |0\rangle_j
     \label{eq:init_singlet_gate}
     \end{equation}
     with 
     \begin{equation}
        \hat{U}^{s}_{ij} = \hat{CX}_{ij} \hat{H}_{i} \hat{X}_{j} \hat{X}_{i},
    \end{equation}
    where 
    $\hat{X}_{i}$ ($\hat{H}_{i}$) is the Pauli X (Hadamard) gate at qubit $i$ 
    and $\hat{CX}_{ij}$ is the controlled-NOT gate with qubit $i$ being the control qubit, resulting in 
    $|s\rangle_{i,j} =\frac{1}{\sqrt{2}}(|0\rangle_i|1\rangle_j-|1\rangle_i|0\rangle_j)$. The quantum circuits for these six initialization 
    layers in Eqs.~(\ref{eq:init_lays_s})--(\ref{eq:init_lays_d}) are explicitly shown in Fig.~\ref{fig:schm_fig}.

    Note that the proper basic entanglement structure can be established by the corresponding initialization layers for specific phases, 
    because $|\phi_s\rangle$, $|\phi_{00}\rangle$, $|\phi_{01}\rangle$, $|\phi_{10}\rangle$, and $|\phi_{11}\rangle$ 
    in Eqs.~(\ref{eq:init_lays_s})--(\ref{eq:init_lays_11}) are all the fixed point state for the SPT Haldane phase, 
    and $|\phi_d\rangle$ in Eq.~(\ref{eq:init_lays_d}) is the fixed point state for the trivial dimer phase. Indeed, as shown in 
    Appendix~\ref{app:entanglement}, these circuit states already exhibit the degeneracy of 
    entanglement spectrum expected for these two phases.
    We should also emphasize that this can be achieved with the depth of the circuit that does not scale with the system size, 
    although the number of gates apparently increases linearly with the system size. 
    Notice also that the four edge mode patterns in the SPT Haldane 
    phase can be engineered by implementing their precursors in the initialization layer, 
    as in $|\phi_{00}\rangle$, $|\phi_{01}\rangle$, $|\phi_{10}\rangle$, and $|\phi_{11}\rangle$, which are mutually orthogonal.

\begin{figure}[hbtp]
        \centering
        \includegraphics[width=\linewidth]{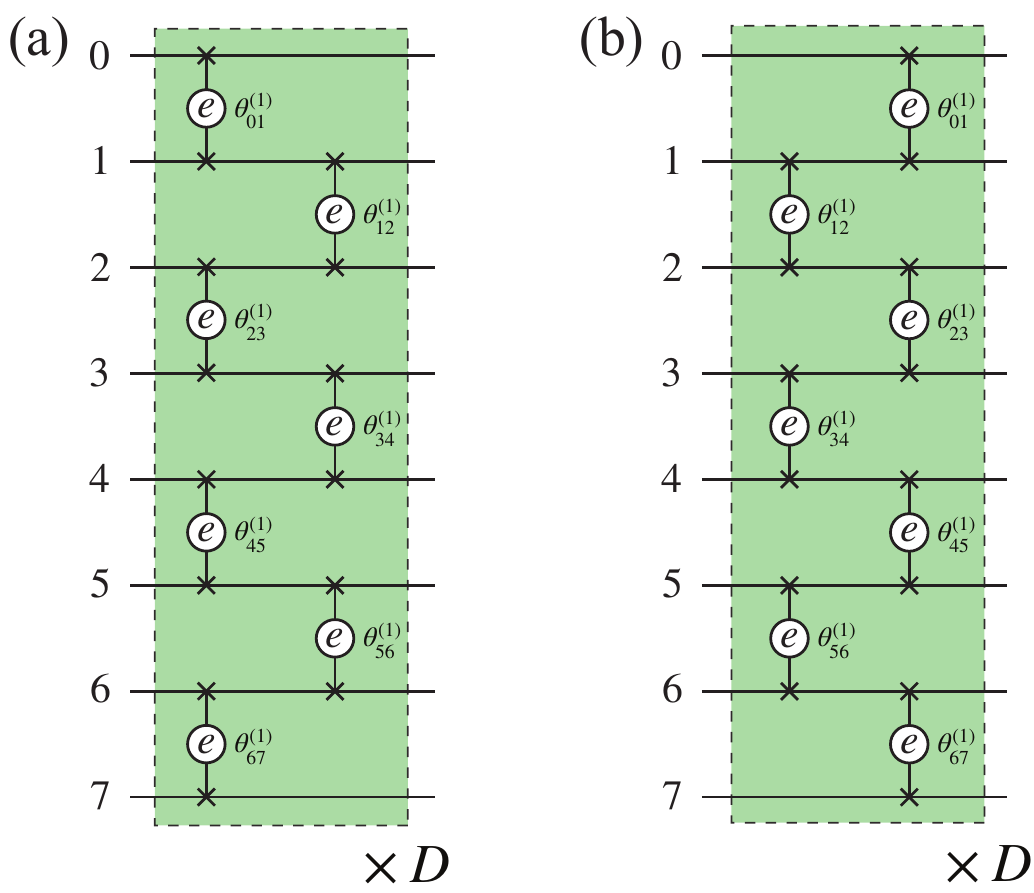}
        \caption{\label{fig:eswap}
            The first unit of the variational layer $\hat{C}(\boldsymbol{\theta})$ for (a) the first five initialization layers in 
            Fig.~\ref{fig:schm_fig} and (b) the last initialization layer in Fig.~\ref{fig:schm_fig}. 
            Here, we assume $L=8$ and the elementary unit containing 7 eSWAP gates (denoted by ``$e$")
            is repeated $D$ times in the variational layer. Note that the variational parameters $\boldsymbol{\theta}$ in 
            different eSWAP gates, given by $\theta^{(d)}_{ij}$ for the eSWAP gate acting on qubits $i$ and $j$ in the $d$-th 
            unit ($d=1,2,\cdots,D$), are treated as independent parameters.}
    \end{figure}

    For the variational layer, we consider a group of brick-wised eSWAP 
    gates~\cite{loss1998quantum,divincenzo2000universal,brunner2011two,lloyd2014quantum,lau2016universal,seki2020symmetry}, 
    which is repeated $D$ times, hence the depth of the circuit \textit{Ansatz} being represented by $D$. 
    As shown in Fig.~\ref{fig:eswap}, the explicit arrangement of eSWAP gates depends on the initialization layer, 
    acting them first on the pairs of qubits 
    in which the singlets are not formed in the initialization layer. 
    A single eSWAP gate acting at qubits $i$ and $j$ is defined by the two-qubit unitary $\hat{U}_{ij}(\theta)$ as \cite{seki2020symmetry}
    \begin{eqnarray}
        \label{eq:eswap_gate}
        \nonumber
        \hat{U}_{ij}(\theta) &=& \exp(-i\theta \hat{P}_{ij}/2) \\
        &=& \cos(\theta/2) \hat{I} - i\sin(\theta / 2)\hat{P}_{ij}~,
    \end{eqnarray}
    where $\hat{P}_{ij}$ is the SWAP gate acting at qubits $i$ and $j$, and $\theta$ is a variational parameter assumed real. 
    Therefore, each 
    eSWAP gate contains a single variational parameter, denoted by $\theta^{(d)}_{ij}$ for the eSWAP gate acting at qubits 
    $i$ and $j$ in the $d$-th unit of the variational layer $\hat{C}(\boldsymbol{\theta})$ (see Fig.~\ref{fig:eswap}), 
    and these variational parameters $\boldsymbol{\theta}$ are treated as independent variables.

    The physical insight of employing the eSWAP gate comes from its relation to the resonating valence bond (RVB) 
    state~\cite{pauling1933calculation,anderson1973resonating}, known as a very faithful variational state to approximate various 
    many-body ground states of quantum spin systems~\cite{becca2017quantum}. A series of eSWAP gates acting on a particular direct 
    product of many singlets which covers the lattice, i.e., a valence bond solid (VBS) state, can mix up these singlets to create 
    an RVB state, although the coefficients of each valence bond are correlated in someway~\cite{seki2020symmetry}. 
    For the simplest example, considering one eSWAP gate acting on two singlets, we have
    \begin{equation}
        \label{eq:eswap_gen_rvb}
        \hat{U}_{jk}(\theta)|s\rangle_{i,j}|s\rangle_{k,l} = \cos(\theta/2)|s\rangle_{i,j}|s\rangle_{k,l} - i\sin(\theta/2)|s\rangle_{i,k}|s\rangle_{j,l},
    \end{equation}
    thus generating different two singlets.  
    The eSWAP circuit \textit{Ansatz} has been applied to the VQE calculations of the one-dimensional (1D) spin-1/2 isotropic 
    antiferromagnetic Heisenberg model and obtained very accurate results~\cite{seki2020symmetry}.

    Having constructed the explicit form of the variational quantum circuit \textit{Ansatz}, we shall now discuss its symmetry properties. 
    For the initialization layer, its building block, $\hat{U}^{s}_{ij}$, breaks the time-reversal symmetry and the $Z_{2} \times Z_{2}$ symmetry 
    with respect to $\pi$ rotations referring to a pair of orthogonal axes. Accordingly, $\hat{\mathcal{C}}_{0}$ also breaks these two 
    symmetries. 
    When we consider the inversion, which refers to the bond (connecting neighboring unitcells) center, the $\hat{U}^{s}_{ij}$ located on that 
    bond will change the sign, i.e., $\hat{U}^{s}_{ij} = -\hat{U}^{s}_{ji}$ with $i=L/2-1$ and $j=L/2$. 
    Therefore, $\hat{\mathcal{C}}_{0}$ in $|\phi_{00}\rangle$, $|\phi_{01}\rangle$, $|\phi_{10}\rangle$, and $|\phi_{11}\rangle$ 
    also breaks the bond central inversion symmetry (i.e., odd parity), while $\hat{\mathcal{C}}_{0}$ in $|\phi_d\rangle$ 
    preserve this symmetry (i.e., even parity), 
    assuming that $L$ is a multiple of 4~\cite{noteparity}.

    Considering now the variational layer, it is straightforward to verify that $\hat{U}_{ij}(\theta)$ preserves the bond central inversion 
    symmetry, 
    i.e., $\hat{U}_{ij}(\theta_{ij}^{(d)}) = \hat{U}_{ji}(\theta_{ji}^{(d)})$ with $j=L-1-i$, provided that $\theta_{ij}^{(d)}=\theta_{ji}^{(d)}$, 
    and the $Z_{2} \times Z_{2}$ symmetry with respect to $\pi$ rotations referring to 
    a pair of orthogonal axes, i.e., $\hat{U}_{ij}(\theta_{ij}^{(d)}) \overset{Z_{2}\times Z_{2}}{\longrightarrow} \hat{U}_{ij}(\theta_{ij}^{(d)})$, 
    but breaks the time reversal symmetry, i.e., 
    $\hat{U}_{ij}(\theta_{ij}^{(d)}) \overset{\mathcal{T}}{\longrightarrow} \hat{U}_{ij}(-\theta_{ij}^{(d)})$. Consequently, 
    as the product of a serial of eSWAP gates, the variational layer preserves the $Z_{2}\times Z_{2}$ symmetry (for any case) 
    and the bond central inversion symmetry (for the case with reflectable parameters $\{\theta_{ij}^{(d)}\}$) 
    but breaks the time-reversal symmetry. We should note that the entanglement structure implemented in the initialization layer 
    is not destroyed by these eSWAP 
    gates in the variational layer, regardless of the values of variational parameters $\boldsymbol{\theta}$, provided that the number 
    $D$ of layers is not large enough to propagate the quantum information throughout the whole system, 
    i.e., $D<L/4$ (see Appendix~\ref{app:entanglement}).

    Finally, it is instructive to explicitly give an example showing that the SPT state can indeed be connected to a trivial state 
    by a finite depth quantum 
    circuit that breaks certain symmetries. For instance, we can construct a specific circuit $\hat{\mathcal{C}}'$, which evolves 
    the SPT state $|\phi_{00}\rangle$ to the trivial dimer state $|\phi_d\rangle$, i.e., 
    \begin{equation}
    |\phi_d\rangle = \hat{\mathcal{C}}'|\phi_{00}\rangle
        \label{eq:evo_00_to_d}    
    \end{equation}
    with
    \begin{equation}
        \hat{\mathcal{C}}' = \hat{\mathcal{C}}^{d}_{0}(\hat{\mathcal{C}}^{00}_{0})^{-1}. 
    \end{equation}
    $\hat{\mathcal{C}}'$ has a finite depth not scaling with the system size and breaks all symmetries 
    which protect the SPT Haldane phase. In contrast, we will numerically demonstrate in Sec.~\ref{sec:expressibility} that 
    a finite depth circuit cannot perform a similar 
    deformation when the circuit preserves at least one of the symmetries protecting the SPT phase 
    (also see Appendix~\ref{app:entanglement}).

    \subsection{Optimization method}
    The variational parameters $\boldsymbol{\theta}$ in $\hat{\mathcal{C}}(\boldsymbol{\theta})$ are optimized by employing 
    the natural gradient descent (NGD) method~\cite{amari1998natural}. The variational parameters at the $k$-th iteration, 
    $\boldsymbol{\theta}_{k}$, are updated from $\boldsymbol{\theta}_{k-1}$ as
    \begin{equation}
        \label{eq:ngd_update}
        \boldsymbol{\theta}_{k} = \boldsymbol{\theta}_{k-1} - \alpha \frac{1}{\text{Re}\boldsymbol{G}(\boldsymbol{\theta}_{k-1})} \boldsymbol{\nabla}E(\boldsymbol{\theta}_{k-1})~,
    \end{equation}
    where $E(\boldsymbol{\theta}) = \left\langle \Psi(\boldsymbol{\theta})|H|\Psi(\boldsymbol{\theta})\right\rangle$ is the variational energy 
    and $\alpha$ is the descent step. $\boldsymbol{G}(\boldsymbol{\theta})$ is the metric tensor \cite{provost1980riemannian} of 
    the parameter space $\boldsymbol{\theta}$ associated with the variational wavefunction $|\Psi(\boldsymbol{\theta})\rangle$, 
    whose matrix element reads as
    \begin{eqnarray}
        \label{eq:metric_ten_elem}\nonumber
        \left[\boldsymbol{G}(\boldsymbol{\theta})\right]_{ij} =&& \langle \partial_{\theta_{i}} \Psi(\boldsymbol{\theta})|\partial_{\theta_{j}} \Psi(\boldsymbol{\theta})\rangle - \\
        && \langle \partial_{\theta_{i}} \Psi(\boldsymbol{\theta})|\Psi(\boldsymbol{\theta})\rangle \langle \Psi(\boldsymbol{\theta})|\partial_{\theta_{j}}\Psi(\boldsymbol{\theta})\rangle~.
    \end{eqnarray}
    Here, $\lvert\partial_{\theta_{i}}\Psi(\boldsymbol{\theta})\rangle$ is the partial derivate of $\left|\Psi(\boldsymbol{\theta})\right\rangle$ 
    with respect to $\theta_{i}$. The NGD optimization has been successfully applied in the previous VQE study of the 1D Heisenberg 
    model~\cite{seki2020symmetry}. Moreover, relevant numerical methods sharing the same idea with the 
    NGD method~\cite{sorella2001generalized,Yunoki2006,haegeman2011time} have been widely used to study quantum 
    many-body systems.

    In the calculation of $\boldsymbol{\nabla}E(\boldsymbol{\theta})$ and $\boldsymbol{G}(\boldsymbol{\theta})$, we need to frequently 
    calculate the partial derivative, $\lvert\partial_{\theta_{i}}\Psi(\boldsymbol{\theta})\rangle$. Here, we employ the parameter-shift 
    rule~\cite{Li2017,guerreschi2017practical,Mitarai2018,Schuld2019,seki2020symmetry}, and apply it as
    \begin{equation}
        \label{eq:para_shift_rule}
        \lvert\partial_{\theta_{i}}\Psi(\boldsymbol{\theta})\rangle = \frac{1}{2}\lvert\Psi(\boldsymbol{\theta} + \pi \hat{\boldsymbol{i}})\rangle~,
    \end{equation}
    where $\hat{\boldsymbol{i}}$ is the unit vector for the $i$th dimension in the parameter space.

    For the numerical simulations in Sec.~\ref{sec:haldane_phase}, we choose the initial variational parameters 
    $\boldsymbol{\theta}_{0} = \boldsymbol{0}$ since we deliberately start the optimization of the variational parameters 
    in the variational layer 
    that is applied onto the fixed-point state constructed in the initialization layer. 
    We fix $\alpha = 0.01$ and perform 1000 steps of optimization to reach the well-converged state.
    
    \section{Exploring SPT Haldane phase\label{sec:haldane_phase}}
    With the help of the circuit \textit{Ansatz} constructed on the basis of the strategy introduced above, 
    we can obtain the ground state of the AHC with high accuracy by the VQE calculations. 
    In this section, we demonstrate this by showing that the SPT Haldane phase in the AHC can be fully represented by the shallowest 
    circuit state with $D=1$ both in numerical simulations and on real quantum devices.

    \subsection{Numerical simulations}\label{sec:numerical}
    We simulate the VQE calculations for the ground state of the $L = 16$ AHC under OBC using the statevector method 
    provided by \textit{Qiskit}~\cite{qiskit}. Furthermore, the ground state and the low-lying excited states are also obtained 
    by performing the exact diagonalization (ED) calculations. 
    In the following simulations, we always adopt $|\phi_s\rangle$ as the circuit state in the initialization layer, 
    unless otherwise stated.

    \begin{figure}[h]
        \centering
        \includegraphics[width=\linewidth]{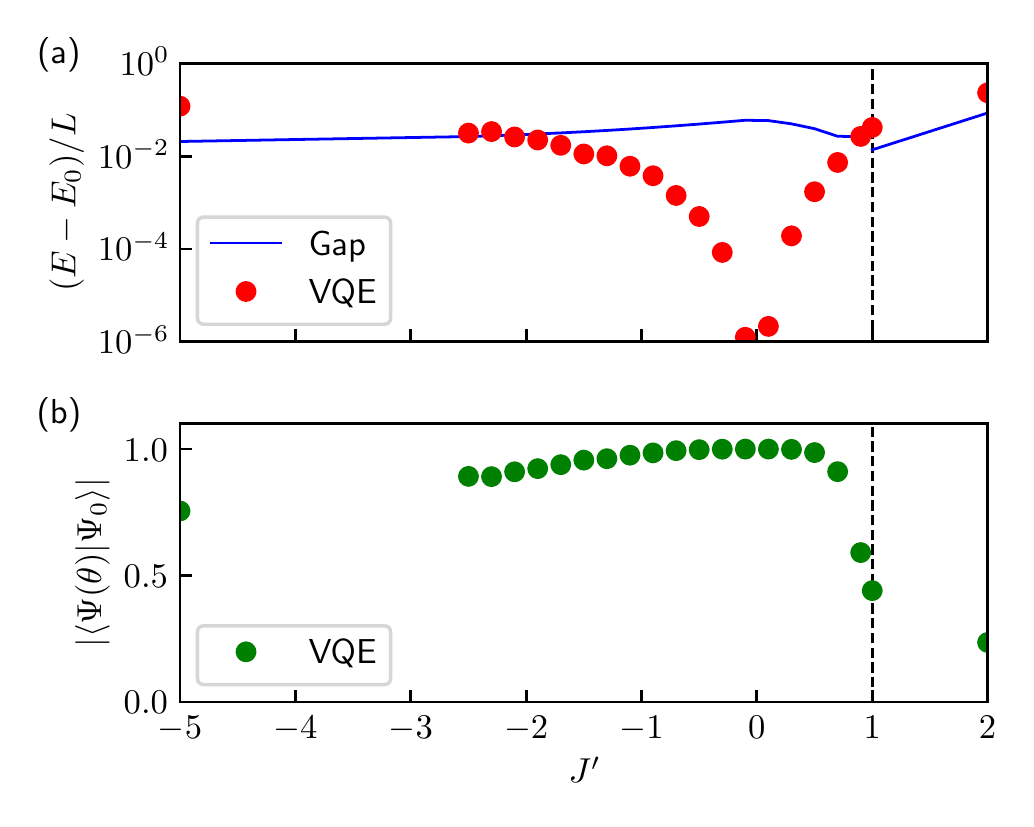}
        \caption{\label{fig:vqe_eng_ovlp_jps}
            (a) Energy difference per site and (b) wavefunction fidelity $F$ between the ground state $|\Psi_0\rangle$ 
            of the $L=16$ AHC under OBC 
            and the corresponding VQE optimized circuit state $|\Psi(\boldsymbol{\theta})\rangle$ with $D = 1$. 
            In (a), the blue solid line represents the Haldane (trivial) gap for 
            $J' < 1$ ($J' > 1$). The Haldane (trivial) gap is estimated by the ED method as the excitation energy to the fourth (first) excited state. 
            The dashed line highlights the critical coupling $J' = 1$. 
            $E_0$ is the exact ground-state energy, while $E=\langle\Psi(\boldsymbol{\theta})|{H}|\Psi(\boldsymbol{\theta})\rangle$ is the 
            corresponding variational energy.
        }
    \end{figure}

    \subsubsection{Ground state energy and wavefunction fidelity}

    The ground-state energy deviations of the VQE optimized circuit state with $D=1$ from the exact ground state are summarized in 
    Fig.~\ref{fig:vqe_eng_ovlp_jps}(a). For comparison, the energy gap obtained by the ED method is also plotted in the same figure. 
    Here, as the energy gap, 
    we consider for $J' < 1$ the Haldane gap given approximately by the excitation energy to the fourth 
    excited state because of the four-fold ground state degeneracy in the SPT phase. 
    For $J' > 1$, the trivial gap of the system is estimated as the excitation gap to the first excited state. 

    As shown in Fig.~\ref{fig:vqe_eng_ovlp_jps}(a), the energy deviations are smaller than the energy gap for $-2.1 \le J' \le 0.7$, 
    indicating that the circuit \textit{Ansatz} is a suitable variational \textit{Ansatz} for the SPT Haldane phase in a finite parameter region. 
    In particular, the energy deviations are significantly small ($<10^{-3}$) in the region near the $J' = 0$ fixed point.

    In addition to the variational energy, the wavefunction fidelity between the circuit state $|\Psi(\boldsymbol{\theta})\rangle$ 
    and the exact ground state $|\Psi_{0}\rangle$, defined as
    \begin{equation}
        \label{eq:wf}
    F = \left| \left\langle \Psi(\boldsymbol{\theta}) | \Psi_{0}  \right\rangle \right|~,
    \end{equation} 
    is also a crucial evaluation for the circuit \textit{Ansatz}. Compatible with the variational energy results, 
    we find in Fig.~\ref{fig:vqe_eng_ovlp_jps}(b) that the 
    wavefunction fidelity $F$ are always larger than 0.9 for $-2 \le J' \le 0.6$, 
    implying a high quality approximation to the ground state. 
    Note also that the wavefunction fidelity $F$ decreases rapidly when $J'$ approaches further closely to the critical point $J' = 1$  
    and exhibits a very small value $\sim 0.25$ once it crosses the critical point 
    in the trivial dimer phase, where this circuit \textit{Ansatz} fails to describe the ground state.

    \begin{figure}[t]
        \centering
        \includegraphics[width=\linewidth]{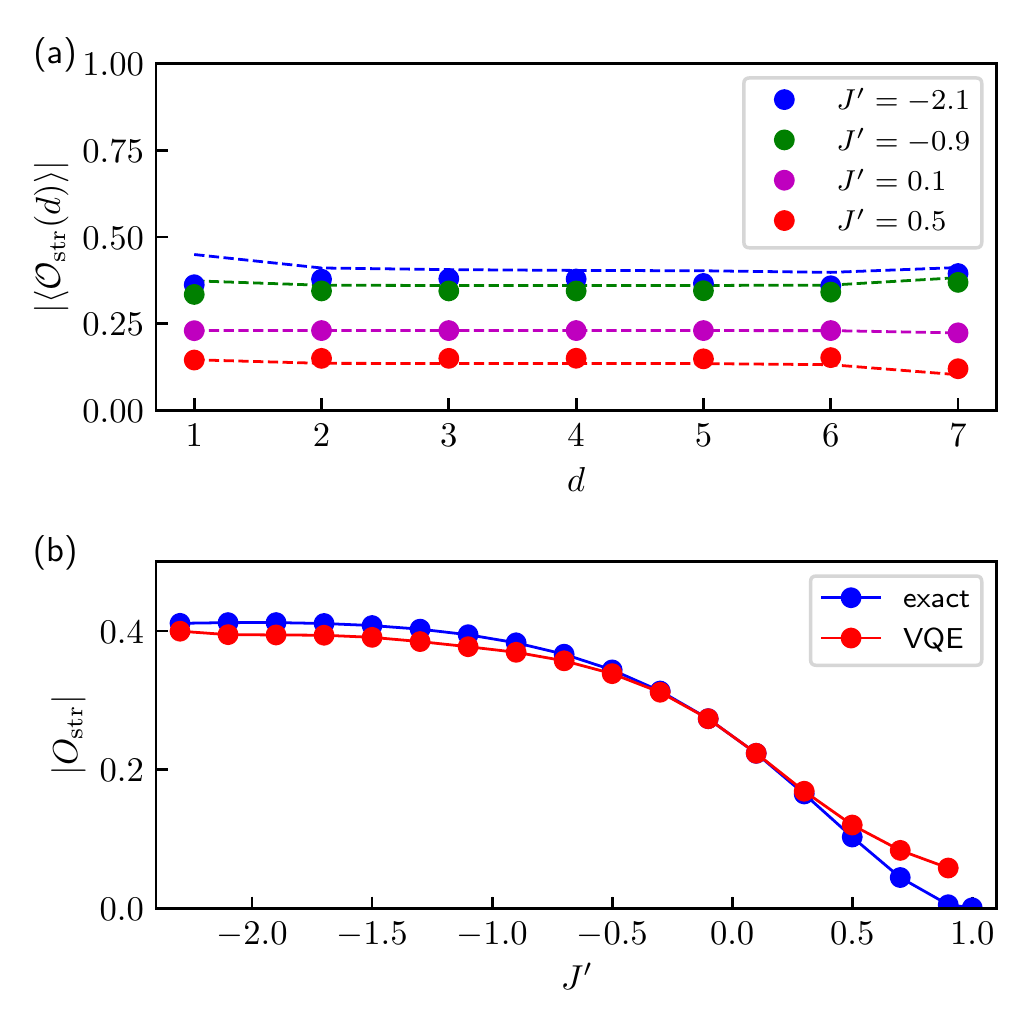}
        \caption{\label{fig:vqe_str_jps}
            String order in the SPT Haldane phase. (a) The expectation value of the string operator, $\langle \mathcal{O}_{\rm str}(d)\rangle$, 
            as a function of distance $d$ for the VQE optimized 
            circuit state $|\Psi(\boldsymbol{\theta})\rangle$ with $D = 1$ (solid circles). 
            For comparison, the corresponding exact results are also shown by dashed lines with the same color. 
            The reference point is fixed at the left edge of the system, i.e., $k = 0$ in Eq.~(\ref{eq:str_op}). 
            (b) The string order parameter $O_{\rm str} = \langle\mathcal{O}_{\rm str}(d=L/2-1)\rangle$ as a function of $J'$ 
            for the VQE optimized circuit states with $D=1$ (red circles) and the exact ground states (blue circles). 
            Here, the system size is $L=16$.}
    \end{figure}

    \subsubsection{String order parameter}
    To measure the nonlocal string operator $\langle \mathcal{O}_{\rm str}(d)\rangle$ in the present finite systems, 
    we fix the reference point to the left end of the system and consider 
    the string order parameter as $O_{\rm str} = \langle\mathcal{O}_{\rm str}(d=L/2-1)\rangle$. 
    Without losing generality, we choose four typical values of $J'$, i.e., $J' = -2.1$, $-0.9$, $0.1$, and $0.5$,  
    in the SPT Haldane phase and display the spatial dependence of the string operator in Fig.~\ref{fig:vqe_str_jps}(a). 
    The results for the VQE optimized circuit states with $D=1$ are in good quantitative agreement with the exact results, 
    although a slight deviation is 
    observed for $J' = -2.1$, which is close to the boundary of the region where the ground state is faithfully expressed by the $D = 1$ 
    circuit \textit{Ansatz} (see Fig.~\ref{fig:vqe_eng_ovlp_jps}). 
    In all of these four different $J'$ values, the string operator does not obviously decay, indicating 
    the long-range string order.

    Figure~\ref{fig:vqe_str_jps}(b) shows the string order parameter $O_{\rm str} $ as a function of $J'$. 
    We find that these results for the VQE optimized circuit state with $D=1$ 
    are also quantitatively compatible with the exact results, except for the region close to the critical point $J'=1$. Therefore, 
    we conclude that the shallow circuit state with $D = 1$ can already represent the string-type correlation rather well.

    \subsubsection{Ground-state degeneracy and edge modes\label{sec:tgt_gss}}

    One of the hallmarks of the SPT Haldane phase is the four-fold degeneracy of the ground state associated with edge modes 
    when the system has open boundaries. Using the eSWAP circuit \textit{Ansatz} with the circuit states $|\phi_{00}\rangle$, 
    $|\phi_{01}\rangle$, $|\phi_{10}\rangle$, and $|\phi_{11}\rangle$ in the initialization layer, 
    we can obtain all these four-fold degenerate ground states 
    in the corresponding optimized circuit states.

    In practice, since $\hat{\mathcal{C}}(\boldsymbol{\theta})|\phi_{00}\rangle$ ($\hat{\mathcal{C}}(\boldsymbol{\theta})|\phi_{01}\rangle$) 
    and $\hat{\mathcal{C}}(\boldsymbol{\theta})|\phi_{11}\rangle$ ($\hat{\mathcal{C}}(\boldsymbol{\theta})|\phi_{10}\rangle$) are 
    related by a global unitary transformation
    \begin{equation}
        \label{eq:ux_trans}
        \hat{U}_{X} = \prod_{i = 0}^{L-1}\hat{X}_{i}~,
    \end{equation}
    i.e., $\hat{U}_X \hat{\mathcal{C}}(\boldsymbol{\theta})|\phi_{00}\rangle = - \hat{\mathcal{C}}(\boldsymbol{\theta})|\phi_{11}\rangle$ and 
    $\hat{U}_X \hat{\mathcal{C}}(\boldsymbol{\theta})|\phi_{01}\rangle = - \hat{\mathcal{C}}(\boldsymbol{\theta})|\phi_{10}\rangle$ for 
    $L$ being a multiple of 4, and 
    $\hat{U}_X H \hat{U}_X = H$, 
    we only perform the VQE simulations for the circuit \textit{Ans\"{a}tze} with $|\phi_{00}\rangle$ and $|\phi_{01}\rangle$ in the 
    initialization layer and reuse the optimized 
    parameters to generate the optimized circuit states with $|\phi_{11}\rangle$ and $|\phi_{10}\rangle$ in the initialization layer.

    As in the case of the string order in Fig.~\ref{fig:vqe_str_jps}, 
    we consider the four typical $J'$ values, i.e., $J' = -2.1$, $-0.9$, $0.1$, and $0.5$, in the SPT Haldane phase. 
    For all these different $J'$ values, 
    we find that the ground-state energies obtained for the VQE optimized circuit states, 
    containing only a $D=1$ layer, with the four different circuit states in the initialization 
    layer are nearly degenerate, and their deviations from the exact ground-state energies are 
    always smaller than the corresponding Haldane gap. Furthermore, these four states are orthogonal with each other; 
    the VQE optimized circuit states 
    $\hat{\mathcal{C}}(\boldsymbol{\theta})|\phi_{00}\rangle$ ($\hat{\mathcal{C}}(\boldsymbol{\theta})|\phi_{01}\rangle$) and 
    $\hat{\mathcal{C}}(\boldsymbol{\theta})|\phi_{11}\rangle$ ($\hat{\mathcal{C}}(\boldsymbol{\theta})|\phi_{10}\rangle$) are exactly 
    orthogonal by construction and the wavefunction fidelities
    among other circuit states are found numerically to be $<10^{-9}$. 
    Therefore, we have four mutually-orthogonal nearly-degenerate states in the ground state manifold.

\begin{figure}
        \centering
        \includegraphics[width=\linewidth]{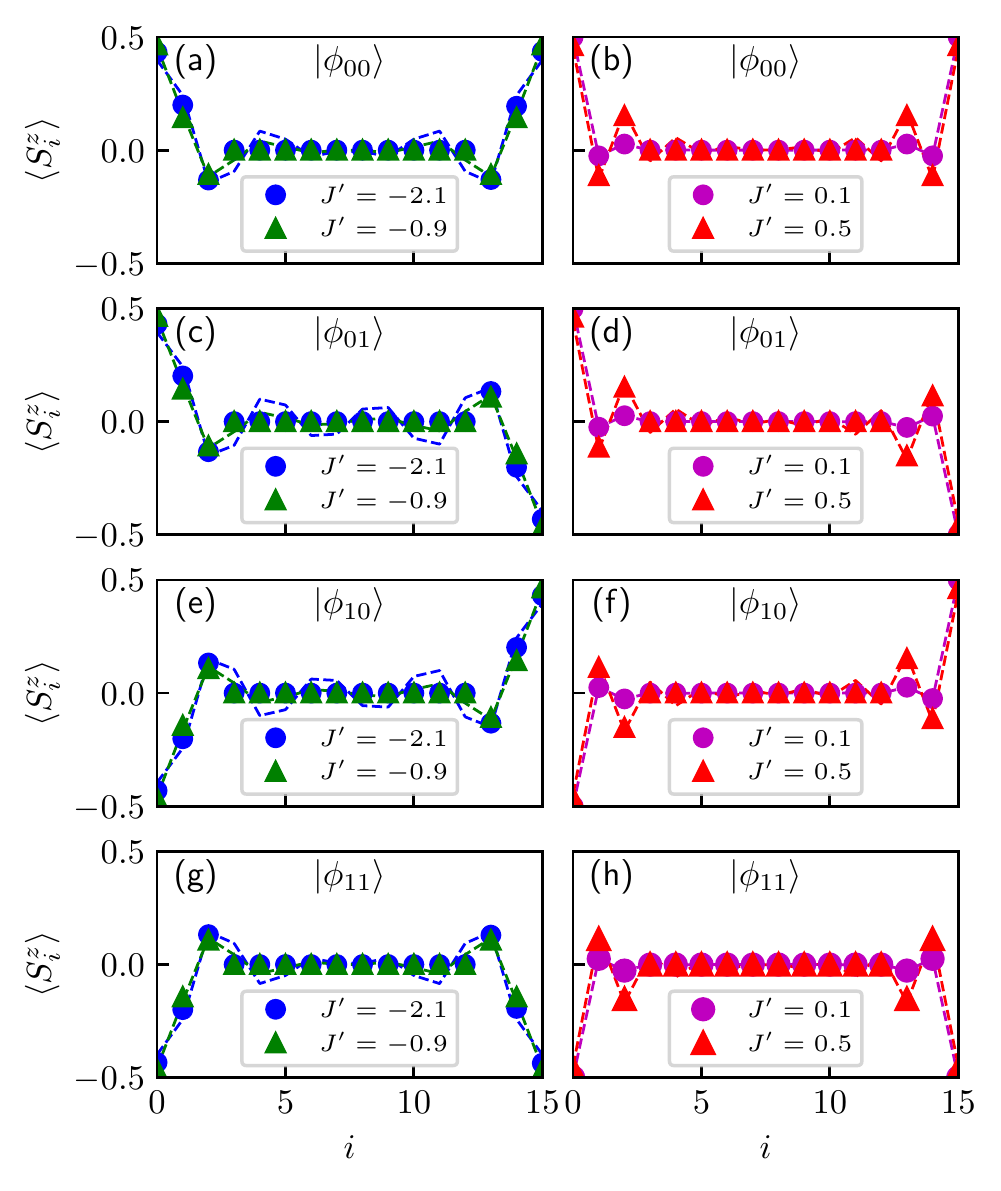}
        \caption{\label{fig:vqe_sz_profile}
            The spatial distribution profile of the on-site magnetization $\left\langle S^{z}_{i} \right\rangle$ 
            for the VQE optimized circuit states containing a $D=1$ layer with different initializations in the SPT Haldane phase at 
            $J'=-2.1$, $-0.9$, 0.1, and 0.5   
            (solid circles and triangles). (a, b) 
            the circuit states $|\phi_{00}\rangle$, (c, d) $|\phi_{01}\rangle$, (e, f)  
            $|\phi_{10}\rangle$, and (g, h) $|\phi_{11}\rangle$ are employed in the initialization layer. 
            For comparison, the exact results are also shown by dashed lines  
            with the same color. 
            Note that although the results in (a) and (b) [(c) and (d)] are exactly related to those in (g) and (h) [(e) and (f)], respectively, 
            because of the construction of the circuit states (see the text) and $\hat{U}_X S_i^z \hat{U}_X = -S_i^z$, 
            here we show these results for completeness.}
    \end{figure}

    To further explore the properties of these four VQE optimized circuit states, we probe the edge mode by evaluating the on-site 
    magnetization $\langle S^{z}_{i} \rangle$. As shown in Fig.~\ref{fig:vqe_sz_profile}, we indeed find distinct edge mode 
    patterns associated with the specific initialization. 
    It is also confirmed that the localization length of the edge mode is shorter when $J'$ is closer to 0. 
    The results for the VQE optimized circuit states are in good agreement with the exact results, although they are slightly 
    deviated from the exact values for $J' = -2.1$, which is  
    close to the expressibility limit of the $D = 1$ variational circuit state (see Fig.~\ref{fig:vqe_eng_ovlp_jps}).
    These results clearly demonstrate that the properly constructed variational quantum circuit state with shallow circuit layers 
    can capture the details of exotic quantum many-body phases, such as the edge modes of an SPT phase 
    by engineering four edge mode patterns in the initialization layer.

    \subsection{Real quantum device demonstration}
    
    To demonstrate that the circuit \textit{Ansatz} is also suitable for the real quantum devices, here the VQE optimized circuit states 
    obtained classically by the numerical simulations are evolved on the real quantum devices. 
    As a demonstration, we consider the $L = 8$ system with $J' = 0.15$ and the VQE optimized circuit states with $D = 1$ 
    for evaluating the string order and the spatial distribution of $\langle S_i^z\rangle$. 
    By performing the quantum tomography for the $L = 4$ system, we also evaluate bipartite entanglement spectra 
    in the SPT Haldane phase at $J'=0.1$ and the trivial dimer phase at $J'=10$  
    using the VQE optimized circuit states with $D = 1$. 
    
\begin{figure}[t]
        \centering
        \includegraphics[width=\linewidth]{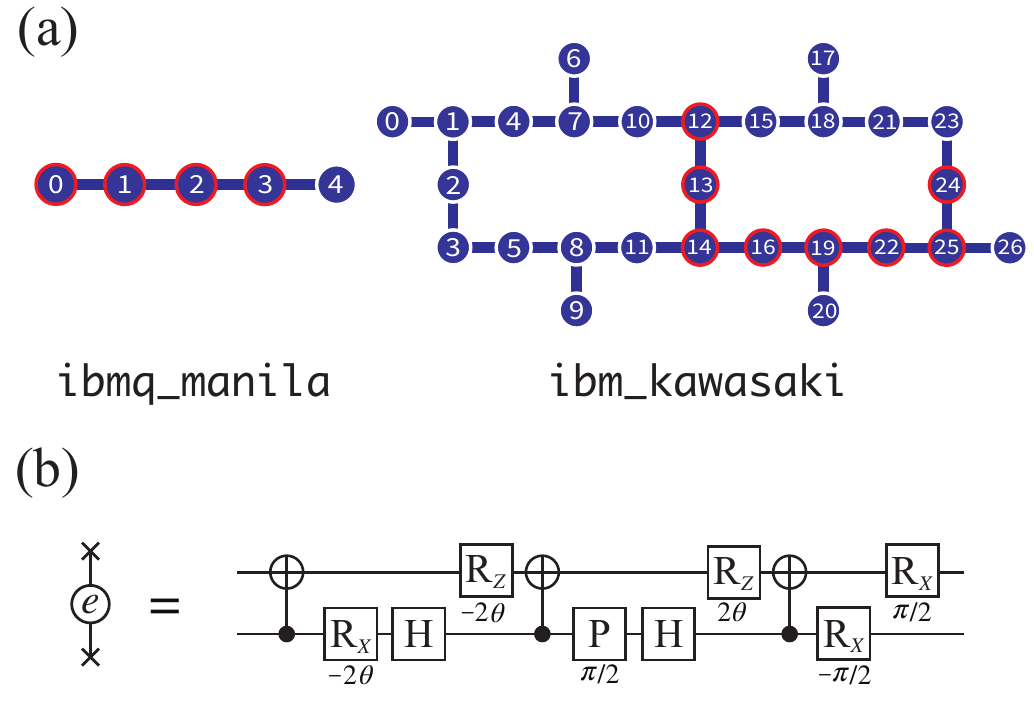}
        \caption{\label{fig:real_device_setup}
        (a) Qubit allocation (marked by red circles) used for the $L = 4$ and $8$ systems in the IBM quantum devices 
        \texttt{ibmq\_manila} (left) and \texttt{ibm\_kawasaki} (right). (b) Decomposition of the eSWAP gate 
        $\exp(-{\rm i}\theta\hat{P}_{ij}/2)$ into single qubits gates and CNOT gates for real quantum device demonstration. 
        Here, ``R$_{X(Z)}$" is the single-qubit rotation gate about the $x(z)$ axis acting on qubit $i$, i.e., 
        $\hat{R}_{X(Z)}(\lambda)=\exp(-i\lambda\hat{X}_i(\hat{Z}_i)/2)$ with the rotation angle $\lambda$ indicated in the figure, and  
        ``P" is the phase gate given as the diagonal matrix $\hat{P}(\lambda)=\rm{diag}[1,e^{i\lambda}]$ with the phase $\lambda$ 
        indicated in the figure.}
    \end{figure}

    For this purpose, we employ the quantum devices (\texttt{ibmq\_manila} and \texttt{ibm\_kawasaki}) provided by IBM 
    Quantum~\cite{ibmq}. 
    The qubit allocations in the quantum devices used for the $L=4$ and $8$ systems are shown in Fig.~\ref{fig:real_device_setup}(a). 
    As shown in Fig.~\ref{fig:real_device_setup}(b), the eSWAP gate defined in Eq.~(\ref{eq:eswap_gate}) is further decomposed 
    into 8 single qubit gates and three CNOT gates in general~\cite{Vidal2004,Chiesa2019} (up to a global phase factor). 
    After the transpilation using {\it Qiskit}, the number of CNOT gates contained 
    in the VQE optimized circuit state with $D=1$ is as many as 24 (10) for the $L=8$ (4) system in total when 
    the circuit state $|\phi_{00}\rangle$, $|\phi_{01}\rangle$, $\phi_{10}\rangle$, or $|\phi_{11}\rangle$ is employed 
    in the initialization layer.
    However, note that additional SWAP gates, which are converted with additional 
    CNOT gates, are required when $|\phi_s\rangle$ is employed in the initialization layer.
    Each experiment runs 8192 shots to collect the data. 
    To estimate the statistical error, the same experiment with 8192 shots is repeated 10 times.

    \subsubsection{String order}
    
    We should first notice that because of $e^{i\pi S^{z}_j} = i\hat{Z}_j$, the expectation value of the string operator $\mathcal{O}_{\rm str}(d)$ in Eq.~(\ref{eq:str_op}) 
    can be evaluated simply by measuring all the bit strings in the computational basis on the quantum device.  
    Furthermore, since the $z$ component of the total spin, $S^{z}_{\rm tot} = \sum_{j} S^{z}_{j}$, is conserved in the circuit \textit{Ansatz}, 
    the ideal measurement of each set of all bit strings in the computational basis must also conserve the total spin $S^z_{\rm tot}$ at 
    every single experiment. This implies that 
    the real device noise can be mitigated in each experiment by post-selecting only those shots which respect 
    $S^{z}_{\rm tot} = 0$ (or any particular value)~\cite{tan2021realizing}. 
    For comparison, we also evaluate the string order in the trivial dimer phase at $J' = 10$ 
    using the VQE optimized circuit state containing a $D = 1$ layer with $|\phi_d\rangle$ in the initialization layer.

\begin{figure}[t]
        \centering
        \includegraphics[width=\linewidth]{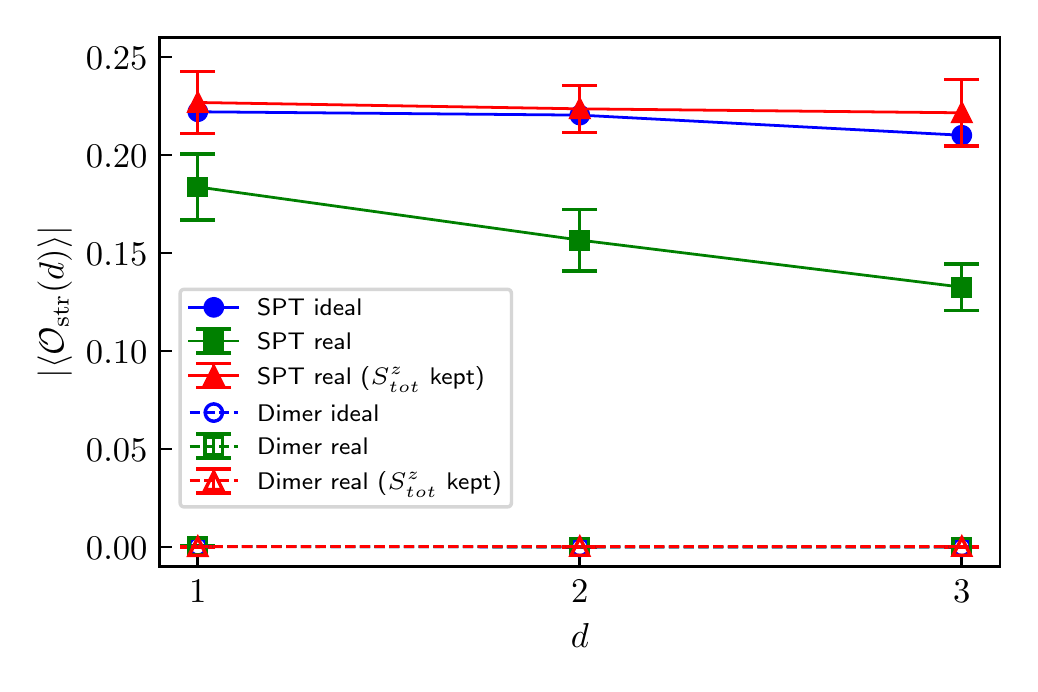}
        \caption{\label{fig:circ_smpl_str}
            The expectation value of string operator, $\langle \mathcal{O}_{\rm str}(d)\rangle$, 
            as a function of distance $d$ in the SPT Haldane phase at $J'=0.15$ 
            (green solid squares) 
            and the trivial dimer phase at $J'=10$ (green open squares) evaluated on the IBM quantum device (\texttt{ibm\_kawasaki}). 
            Here, the system size is $L=8$ and the VQE optimized circuit state containing a $D=1$ layer with $|\phi_s\rangle$ ($|\phi_d\rangle$) 
            in the initialization is used for the SPT Haldane (trivial dimer) phase.
            The error mitigated results by post-selecting shots 
            only satisfying $S^z_{\rm tot}=0$ are also plotted by triangles. 
            The error bars are estimated as the standard error of the mean from 10 independent sets of 
            the same experiments with each having 8192 shots. 
            For comparison, the numerical results obtained for the same VQE optimized circuit states (indicated as ``ideal") 
            are also shown by blue circles. 
            Note that the three results for the trivial dimer phase are indistinguishable in this scale.}
    \end{figure}

    As shown in Fig.~\ref{fig:circ_smpl_str}, even in the small system with $L = 8$, the distinction 
    between the SPT Haldane phase and the trivial dimer phase can be 
    clearly identified by measuring the string operator.
    In the real device experiments, the string operator shows a smaller value with a slightly faster decay as compared with the numerical 
    one in the SPT Haldane phase due to the noise, while it is close to zero in the trivial dimer phase. 
    After post-selecting the shots respecting $S^{z}_{\rm tot} = 0$, the string operator becomes flatter in the SPT Haldane phase,
    clearly demonstrating that these two circuit states can be distinguished on the real quantum hardware.

    \subsubsection{Edge modes}
    
\begin{figure}[ht]
        \centering
       \includegraphics[width=\linewidth]{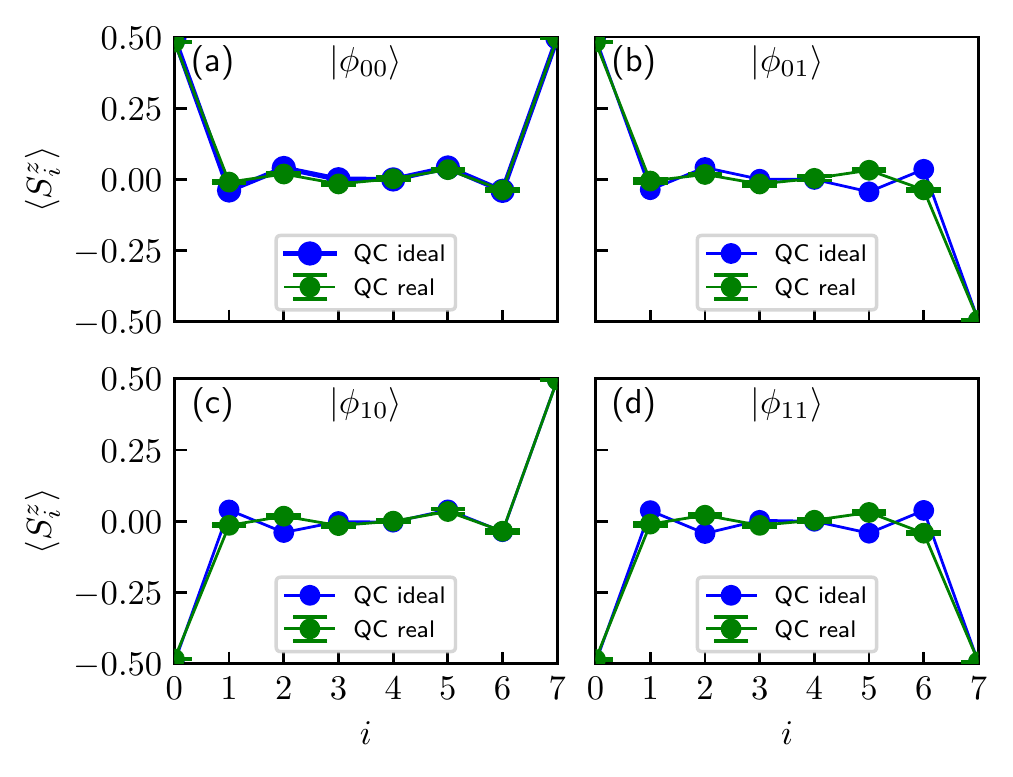}
        \caption{\label{fig:circ_smpl_szprofile}
            The spatial distribution profile of the on-site magnetization $\langle S^{z}_{i}\rangle$ in the SPT Haldane phase at $J' = 0.15$ 
            for the four-fold nearly 
            degenerate ground states represented by the VQE optimized circuit \textit{Ans\"{a}tze} containing a $D=1$ layer 
            with the four different circuit states, 
            (a) $|\phi_{00}\rangle$, (b) $|\phi_{01}\rangle$, (c) $|\phi_{10}\rangle$, and (d) $|\phi_{11}\rangle$, in the initialization layer 
            (green solid circles). 
            The error bars are estimated as the standard error of the mean from 10 independent sets of 
            the same experiments with each having 8192 shots. 
            For comparison, the corresponding numerical results obtained for the same VQE optimized circuit states (indicated as ``ideal") 
            are also shown by blue solid circles.}
    \end{figure}

    Similar to the numerical simulations in Sec.~\ref{sec:tgt_gss}, the four-fold nearly degenerate ground states can also be 
    generated on the real quantum device by constructing the VQE optimized circuit states containing a $D=1$ layer 
    with the four different circuit states, 
    $|\phi_{00}\rangle$, $|\phi_{01}\rangle$, $|\phi_{10}\rangle$, and $|\phi_{11}\rangle$, in the initialization layer. 
    Here, we evolve these four optimized circuit \textit{Ans\"{a}tze} on the real device and evaluate the spatial distribution profile of 
    $\langle S^{z}_{i}\rangle$ by a series of measurements of individual qubits in the computational basis. 
    As shown in Fig.~\ref{fig:circ_smpl_szprofile}, the profiles of $\langle S^{z}_{i}\rangle$ exhibit four clearly distinguished patterns, 
    depending on the different settings in the initialization layer, and they are also consistent with the numerical results obtained 
    for the same VQE optimized circuit states.

    \subsubsection{Entanglement spectrum}
    
    The SPT phase can also be characterized against the trivial product state by exploring the degeneracy of the entanglement spectrum
    \begin{equation}
        \label{eq:es}
        \xi(i) = -\ln(\lambda_{i})~,
    \end{equation}
    where $\lambda_{i}$ is the $i$-th largest eigenvalue of the reduced density matrix for the ground state by tracing out half of 
    the degrees of freedom in the real space of the system, and hence $\xi(i)$ is ordered in the ascent order. 
    The entanglement spectra of the SPT Haldane phase exhibit at least 
    two-fold degeneracy for each spectrum level~\cite{pollmann2010entanglement}. In contrast, this degeneracy is split generally 
    in the trivial dimer phase.

\begin{figure}[ht]
        \centering
        \includegraphics[width=\linewidth]{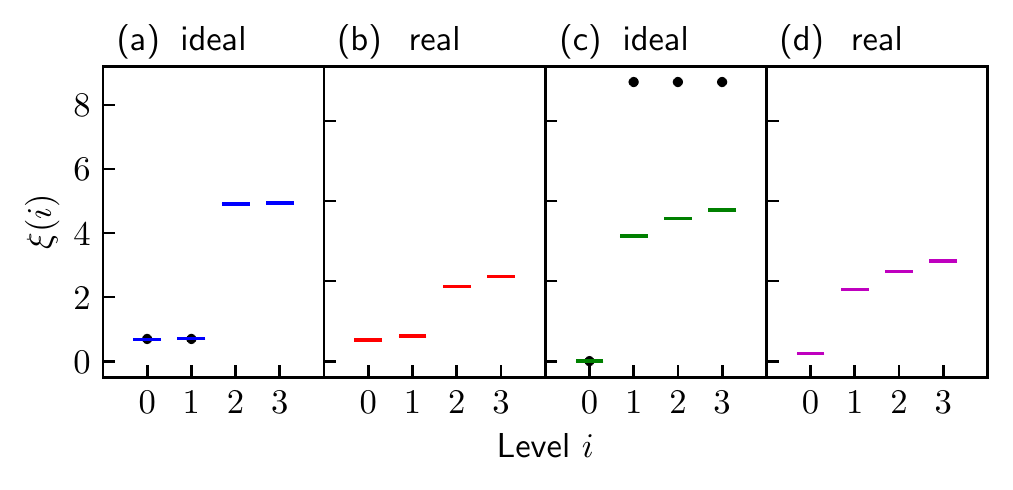}
        \caption{\label{fig:circ_smpl_tomo_es}
            Entanglement spectrum $\xi(i)$ referring to left half of the system with $L = 4$ in (a,b) the SPT Haldane phase at $J'=0.1$ 
            and (c,d) the trivial dimer phase at $J'=10$ evaluated from the VQE optimized circuit states containing a $D = 1$ layer with 
            (a,b) $|\phi_{00}\rangle$ 
             and (c,d) $|\phi_d\rangle$ in the initialization layer. 
             The results in (a) and (c) are calculated by numerical simulations and those in (b) and (d) are evaluated on the IBM quantum 
             device (\texttt{ibmq\_manila}) where the device noise is mitigated by post-selecting only the real part of the reduced density matrix. 
             For comparison, the corresponding exact results are also shown in (a) and (c) by black dots. 
             Note that the exact results for $i=2$ and 3 in (a) are located larger than this scale 
             but they are confirmed numerically to be doubly degenerate.}
    \end{figure}

    The reduced density matrix for a circuit state can be evaluated by quantum tomography~\cite{nielsen2010quantum}. 
    Since a direct procedure of the quantum tomography involves the measurements of all Pauli strings referring to the target 
    segment of the system, the system size which can be treated on the real device is strongly limited. Therefore, we only consider 
    the $L = 4$ system. 
    Here, we evolve the two VQE optimized circuit \textit{Ans\"{a}tze} with $D=1$, which refer to the SPT Haldane phase at $J' = 0.1$ 
    (having $|\phi_{00}\rangle$ in the initialization layer) 
    and the trivial dimer phase at $J' = 10$ (having $|\phi_d\rangle$ in the initialization layer), on the real device 
    and perform the quantum tomography to obtain the reduced density matrices. 
    Figures~\ref{fig:circ_smpl_tomo_es}(b) and \ref{fig:circ_smpl_tomo_es}(d) show the obtained entanglement spectra 
    for the SPT Haldane phase and the trivial dimer phase, respectively. 
    For comparison, the corresponding numerical results for the same VQE optimized circuit states as well as the exact results
    are also shown in Figs.~\ref{fig:circ_smpl_tomo_es}(a) and \ref{fig:circ_smpl_tomo_es}(c).

    We observe that although the spectrum levels slightly deviate from the numerical results, these two circuit states exhibit 
    a distinguishable entanglement spectrum even on the real device. Namely, the first and second spectrum levels are 
    quasi-degenerate in the SPT Haldane phase 
    [see Fig.~\ref{fig:circ_smpl_tomo_es}(b)], while  
    only a single leading spectrum level is observed in the trivial dimer phase 
    [see Fig.~\ref{fig:circ_smpl_tomo_es}(d)], 
    indicating that the state in the latter is close to a direct-product state. Based on these results, 
    we confirm that the SPT Haldane phase hosted in the AHC can be fully captured by the shallowest circuit states 
    on a real quantum device.

    \section{Expressibility of circuit Ansatz \label{sec:expressibility}}
    
    In the previous section, we have demonstrated that a shallow circuit \textit{Ansatz} with $D=1$ can already correctly describe 
    the details of the characteristic features of the SPT Haldane phase in a finite range of the model parameter $J'$. 
    Here, in this section, we discuss the expressibility of this circuit \textit{Ansatz} and in particular show that, as in the definition of 
    a SPT phase,  
    a deeper circuit \textit{Ansatz} is indeed required to connect the SPT Haldane phase to the trivial dimer phase if the \textit{Ansatz} 
    preserves the symmetry protecting the SPT phase. 
    We also show that the expressibility (i.e., computational capacity) of the fixed depth circuit \textit{Ansatz} 
    with the appropriate initialization is insensitive to the system size and scales with the spin correlation length. 
    Therefore, the scalable VQE calculations for a gapped topological phase can be achieved once 
    the circuit representation of a fixed-point state in the gapped topological phase is first correctly constructed in the initialization layer.

    \subsection{Connecting the Haldane phase to the dimer phase}
    
    In order to test how deep a circuit is necessary to connect an SPT state to a trivial state, we perform numerically the VQE calculations 
    for the $L = 12$ system in the trivial dimer phase at $J' = 5$ employing the circuit \textit{Ansatz} with $|\phi_s\rangle$, i.e., 
    a fixed-point state of the 
    SPT Haldane phase, in the initialization layer. Since the variational 
    layer preserves the symmetry which protects the SPT Haldane phase, we expect that a trivial state cannot be obtained in the 
    optimized circuit when its depth is too shallow. Here, we employ the degeneracy of the entanglement spectrum to differentiate 
    the SPT Haldane phase from the trivial dimer phase.

    \begin{figure}[ht]
        \centering
        \includegraphics[width=\linewidth]{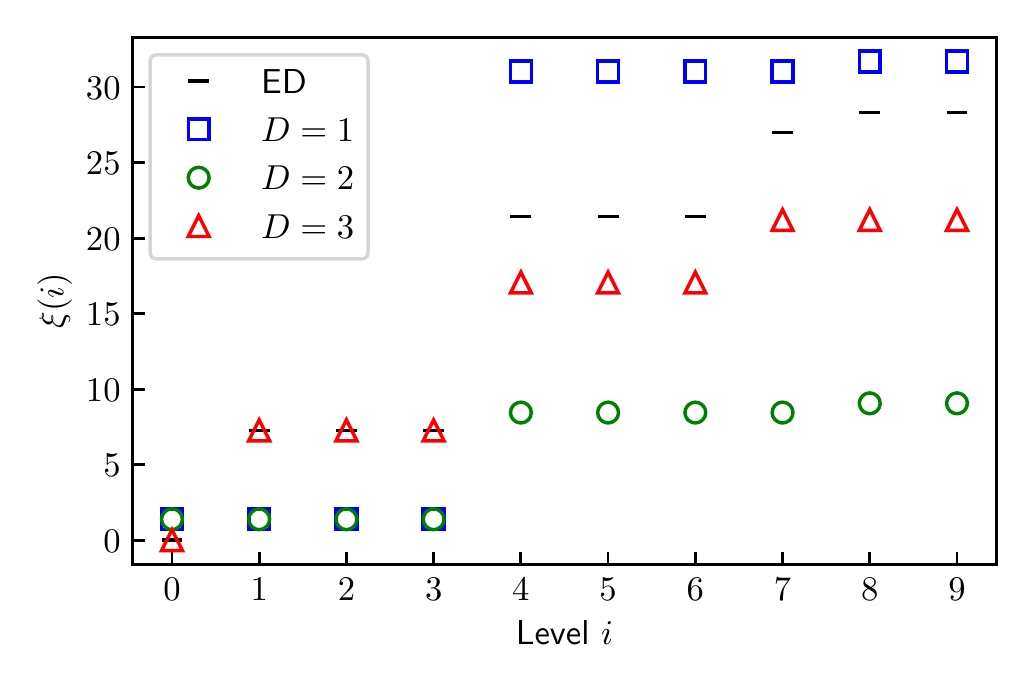}
        \caption{\label{fig:vqe_dimer_phase_es_with_qc_depth}
            The first 10 entanglement spectra $\xi(i)$ of the VQE optimized circuit \textit{Ans\"{a}tze} with $D = 1$ (squares), $2$ (circles), 
            and $3$ (triangles) in the trivial dimer phase at $J'=5$ for $L = 12$. Here, the circuit state $|\phi_s\rangle$, i.e., a fixed-point 
            state of the SPT Haldane phase, is used in the initialization layer. 
            For comparison, the exact results obtained by the ED method are also shown.}
    \end{figure}

    Figure~\ref{fig:vqe_dimer_phase_es_with_qc_depth} shows the low-lying entanglement spectra evaluated from the reduced density 
    matrix of the VQE optimized circuit states with different depths.  
    Here, the reduced density matrix is obtained by bipartitioning the system exactly at half about the center. 
    We find that the lowest four spectrum levels are always degenerate 
    when the circuits are too shallow ($D \le 2$), indicating these circuit states still host a SPT state. 
    As a sharp difference, 
    the entanglement spectrum of the optimized circuit \textit{Ansatz} with $D = 3$ 
    is highly consistent with the exact result, especially in the lowest-lying part of the entanglement spectrum,
    suggesting that the evolution from an SPT state to a trivial product state is accomplished.  
    Indeed, the number of layers composed of local unitary operators necessary to convert two states with different topological 
    indices should scale with the system size $L$~\cite{zeng2019quantum}, 
    and in this particular system with $L=12$, the $D=3$ layers correspond exactly to the point where the causality cone set 
    by the Lieb-Robinson 
    bound for propagating the information via the local two-qubit gates exceeds the system size~\cite{Shirakawa2021}. More discussion on the degeneracy of entanglement spectrum for the circuit states considered here is found in 
    Appendix~\ref{app:entanglement}.

    \subsection{Expressibility vs. correlation length}
    
    One measure for the expressibility of a particular circuit \textit{Ansatz} for a ground state is to simply monitor the energy deviation 
    of the circuit \textit{Ansatz} 
    with the optimized parameters from the exact ground-state energy. Although the spin correlation length is hard to determine 
    straightforwardly in a small finite system, we can use the spin correlation function at a fixed distance from the edge 
    as a reasonable approximation.

\begin{figure} [ht]
        \centering
        \includegraphics[width=\linewidth]{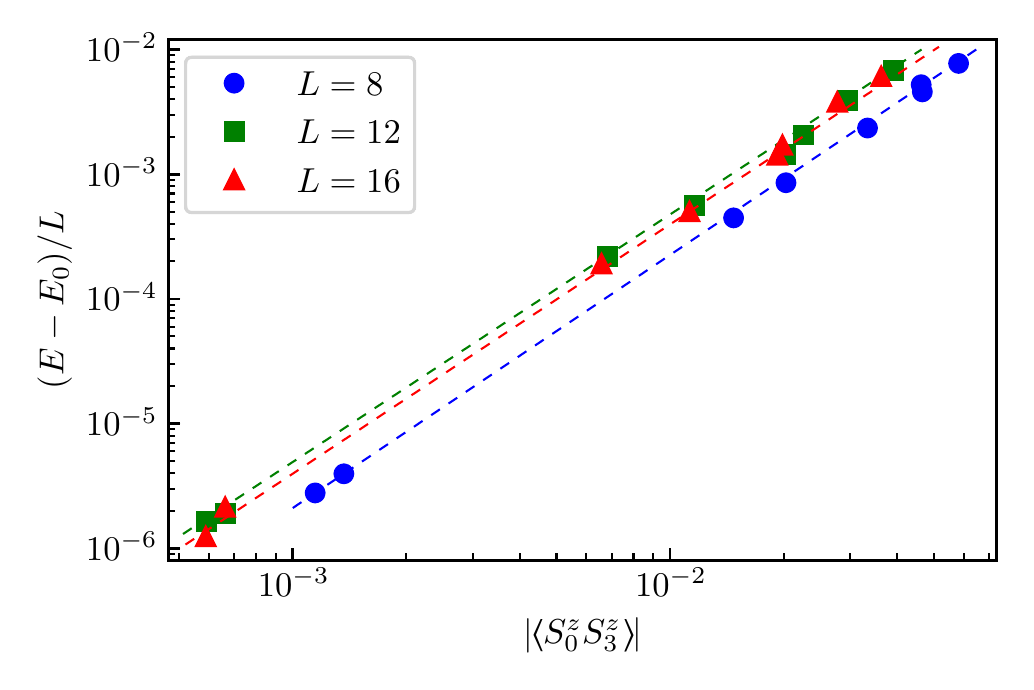}
        \caption{\label{fig:vqe_eng_edg_szsz_r}
            The energy deviation of the VQE optimized circuit \textit{Ansatz} with $D=1$ from the exact ground-state energy $E_0$ 
            versus the spin correlation function at a fixed distance from the edge, $\langle S^z_0 S^z_3\rangle$, 
            evaluated for the exact ground state. The circuit state $|\phi_s\rangle$ is used in the initialization layer of the VQE optimized 
            circuit \textit{Ansatz}. 
            The system sizes considered are $L = 8$ (circles), 12 (squares), and 16 (triangles). 
            The values of $J'$ considered are $-1.1$, $-0.9$, $-0.7$, $-0.5$, $-0.3$, $-0.1$, 0.1, 0.3, and 0.5, 
            which are all in the SPT Haldane phase. The dashed lines are guide for the eye.}
    \end{figure}

    Figure \ref{fig:vqe_eng_edg_szsz_r} shows the relation between the expressibility of the circuit \textit{Ansatz} with $D=1$ 
    and the spin correlation function for the SPT Haldane phase in the AHC.     
    The result clearly reveals that the expressibility for different system sizes has the same scaling behavior with the correlation length,  
    implying that the expressibility of the circuit \textit{Ansatz} with a fixed depth considered here is dictated 
    not by the system size but rather by the correlation length. 
    Moreover, since the energy deviation does not increase with the system size, the computational complexity 
    of the VQE calculations with a given accuracy scales only linearly with the system size (not because of the increase 
    of the circuit depth, but simply because of the increase of the the number of qubits).

    \section{Conclusion and discussion\label{sec:conc_disc}}
    
    In summary, we have proposed a strategy to efficiently construct a variational circuit \textit{Ansatz} for calculating a ground state of a 
    gapped topological phase on NISQ devices. 
    To properly handle both the nontrivial basic entanglement structure contained in the topological 
    phase and the fine structure in the energy landscape depending on the details of the Hamiltonian, 
    an efficient circuit \textit{Ansatz} should be designed in a two-layer structure. In the first layer (i.e., the initialization layer), 
    the basic entanglement structure is constructed by a shallow circuit with a 
    fixed depth that does not scale with the system size 
    (or a circuit with a well-designed structure for an intrinsic topological phase). 
    For the SPT Haldane phase in the AHC, the initialization layer constitutes the ground state of the fixed point 
    Hamiltonian with $J' = 0$. In the second layer (i.e., the variational layer), 
    a parametrized circuit inspired by the Hamiltonian itself is considered to 
    further optimize the state to fit the finite correlations 
    determined by the Hamiltonian parameter away from the fixed point.
    In the AHC, a brick-wised eSWAP 
    circuit is adapted to respect the SU(2) symmetry of the Hamiltonian and preserve one of the symmetries protecting the 
    SPT Haldane phase. 
    All the main features of the SPT Haldane phase, including 
    the long-range string order parameter, the four-fold nearly degenerate ground states corresponding to edge mode patterns 
    in the open system, 
    and the two-fold degenerate entanglement spectrum, have been captured correctly by the optimized circuit state with a very shallow depth 
    ($D=1$) both in numerical simulations and on real quantum devices. 
    Moreover, we have demonstrated that the computational capacity 
    is only sensitive to the correlation length and is independent of the system size. 
    Therefore, the scalable VQE calculation is achieved in this system.

    Although we have only considered the simplest 1D SPT phase in this paper, the circuit \textit{Ansatz} structure proposed here 
    can be directly 
    applied to higher-dimensional cases. Especially for a two-dimensional (2D) SPT phase, whose fixed-point wavefunction is easily 
    constructed by 
    a very shallow circuit~\cite{chen2011two}, a similar variational layer that preserves the symmetry protecting the SPT phase 
    can also be constructed straightforwardly to treat a fine structure of the Hamiltonian. 
    The same idea can also be applied to a gapped topological spin liquid, for which further investigation is however 
    required as to how to generally construct a circuit state for a particular spin liquid phase.

    Moreover, recalling that the depth of the variational layer does not increase with the system size to represent a gapped 
    topological phase with a given accuracy, 
    we realize the potential of 
    achieving the quantum advantage by employing the scheme proposed here. At first glance, the computational complexity 
    for the VQE calculation scales linearly with the system size. However, we should point out that, in principle, 
    all the computation components, such as the calculation of the partial derivative for each parameter 
    and the procedure of energy measurement, can be fully parallelized. In contrast, the computational complexity 
    of the current state-of-the-art tensor network algorithms~\cite{schollwock2011the}, 
    which are considered to be the most efficient classical methods to simulate quantum many-body systems, 
    increases at least linearly with the system size. 
    Therefore, the quantum advantage can be achieved when the system size is very large in one dimension or  
    simply when a 2D system is considered.

    \begin{acknowledgments}
        We acknowledge helpful discussions with Hiroshi Ueda, Shohei Miyakoshi, Takanori Sugimoto, Kazuma Nagao and Hong-Hao Tu. Parts of numerical simulations have been done 
        on the HOKUSAI supercomputer at RIKEN (Project ID No. Q22577) and the supercomputer Fugaku installed 
        in RIKEN R-CCS (Project ID No. hp220217). 
        This work is supported by Grant-in-Aid for Scientific Research (A) (No. JP21H04446) and 
        Grant-in-Aid for Scientific Research (C) (No. JP22K03479) from MEXT, Japan, and JST COI-NEXT (Grant No. JPMJPF2221).
        This work is also supported in part by UTokyo Quantum Initiative, 
        by Program for Promoting Research on the Supercomputer Fugaku (Grant No. JPMXP1020230411) form MEXT, Japan, 
        and by the COE research grant in computational science from 
        Hyogo Prefecture and Kobe City through Foundation for Computational Science.
    \end{acknowledgments}

    \appendix
    
    \section{Entanglement spectrum of the circuit \textit{Ansatz}\label{app:entanglement}}
    
    In this Appendix, we shall examine the degeneracy of the entanglement spectrum extracted from the circuit \textit{Ansatz} 
    proposed in Sec.~\ref{sec:ansatz}. Without losing generality, we fix the system size to $L = 12$ and choose the circuit state 
    $|\phi_{00}\rangle$ in the initialization layer (a fixed-point state for the SPT Haldane phase). 
    For comparison, we also consider 
    a circuit state with $|\phi_{d}\rangle$ in the initialization layer, corresponding to a fixed-point state for the trivial dimer phase. 
    After applying different number $D$ of layers in the variational layer composed of the eSWAP gates with randomly chosen variational 
    parameters $\boldsymbol{\theta}$, 
    we evaluate the entanglement spectrum of the resulting circuit state by numerically diagonalizing the reduced density matrix 
    for the left half of the system. 

    \begin{figure}[ht]
        \centering
        \includegraphics[width=\linewidth]{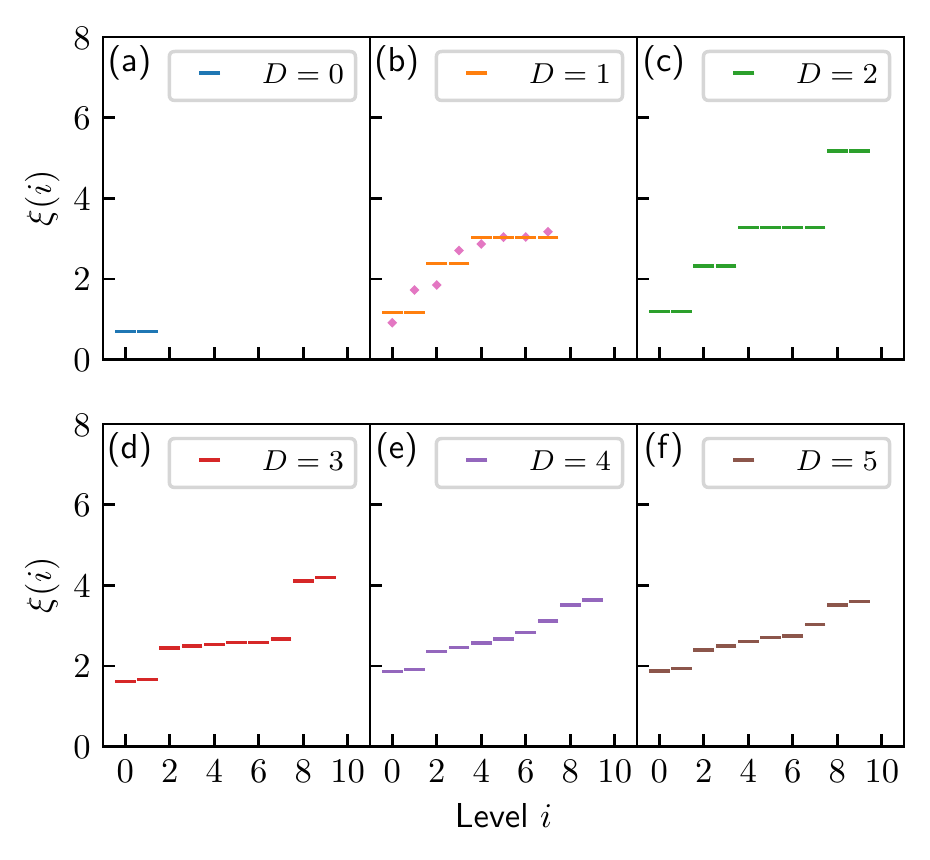}
        \caption{\label{fig:circ_rand_theta_es}
            Low-lying entanglement spectra $\xi(i)$ for the circuit \textit{Ans\"{a}tze} with the circuit state 
            $|\phi_{00}\rangle$ in the initialization layer (corresponding to a fixed-point state for the SPT Haldane phase)
            and varying the depth $D$ in the variational layer composed of the eSWAP gates 
            for which the variational parameters $\boldsymbol{\theta}$ are chosen randomly.  
            The system size is set to $L=12$ and the the entanglement spectrum is evaluated by numerically diagonalizing the reduced 
            density matrix of the system bipartitioned exactly half. 
            Note that the results shown here are obtained for a single set of random parameters, but we have checked that 
            qualitatively the same results are obtained for other sets of random parameters. 
            For comparison, the results for the same circuit state with $D=1$ having the same random variational parameters 
            $\boldsymbol{\theta}$, but containing an additional single CNOT gate acting on qubits 5 and 6 are also plotted 
            by solid diamonds in (b).}
    \end{figure}

    For the circuit states initialized to the fixed-point state of the SPT phase, the entanglement spectrum shows robust 
    (at least) two-fold degeneracy when the depth $D$ of the variational layer is smaller than $L/4$ 
    [see Figs.~\ref{fig:circ_rand_theta_es}(a)--\ref{fig:circ_rand_theta_es}(c)]. Note that the number of nontrivial entanglement 
    spectral levels 
    (i.e., nonzero eigenvalues of the reduced density matrix) increases with $D$ as $2\times4^D$. 
    In sharp contrast, this two-fold degeneracy is lifted entirely once the depth of the variational layer $D \geq L/4$, 
    as shown in 
    Figs.~\ref{fig:circ_rand_theta_es}(d)--\ref{fig:circ_rand_theta_es}(f). This clearly demonstrates that a sufficiently deep circuit is 
    required to transform an SPT state to a trivial state if the circuit respects the symmetry protecting the SPT phase. 
    In order to support this assertion, we also consider exactly the same circuit state with $D=1$ but with an additional single CNOT gate, 
    which breaks the symmetry protecting the SPT phase, acting after the eSWAP gate on qubits 5 and 6 
    [see, e.g., Fig.~\ref{fig:eswap}(a) for $L=8$], 
    and find that only a single CNOT gate is enough to generally lift the 
    two-fold degeneracy of the entanglement spectra, as shown in Fig.~\ref{fig:circ_rand_theta_es}(b) by solid diamonds. 
    We should also note that the circuit depth $D=L/4$ corresponds exactly to the point where the causality cone set 
    by the Lieb-Robinson bound 
    for propagating the information via local two-qubit gates reaches the system size $L$~\cite{Shirakawa2021}. 
    It is also interesting to notice that only the circuit states having $D=L/4$ layers or more can describe all $2^{L/2}$ 
    entanglement spectral levels of the reduced density matrix for a general 1D quantum state when it is bipartitioned exactly half.

    \begin{figure}[ht]
        \centering
        \includegraphics[width=\linewidth]{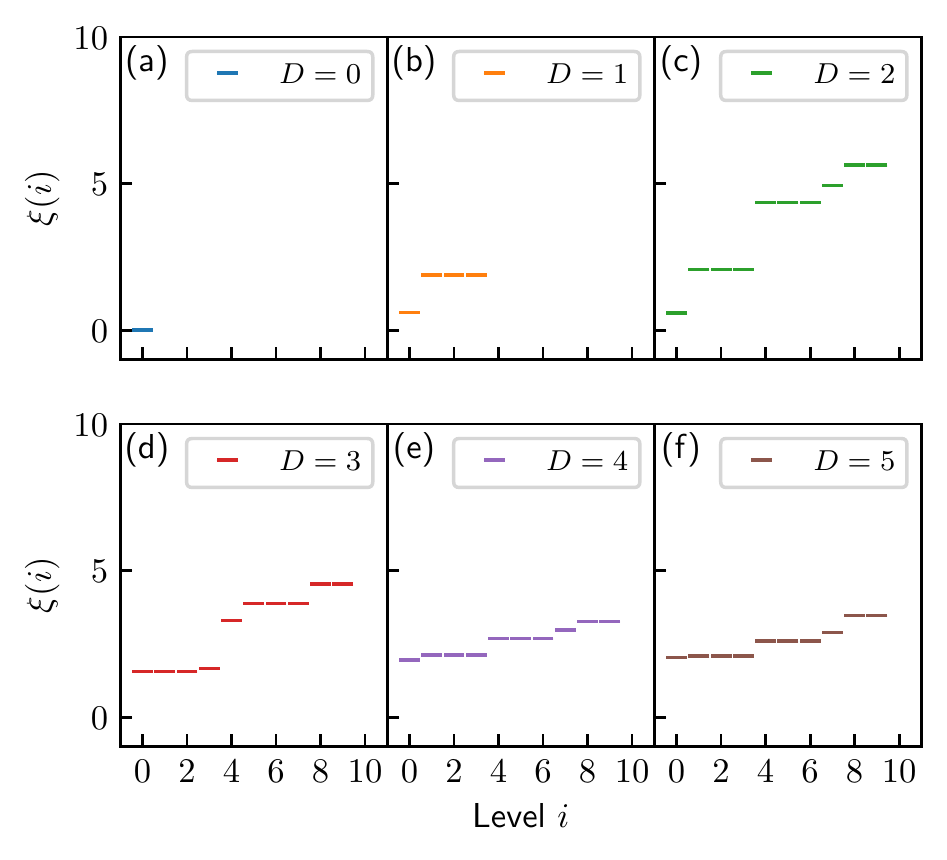}
        \caption{\label{fig:circ_rand_theta_es_d}
            Same as Fig.~\ref{fig:circ_rand_theta_es} but the circuit state $|\phi_{d}\rangle$, 
            corresponding to a fixed-point state for the trivial dimer phase, is adopted in the initialization layer.}
    \end{figure}

    On the other hand, as shown in Fig.~\ref{fig:circ_rand_theta_es_d}, the circuit states initialized to the fixed-point state of the trivial 
    dimer phase exhibit a distinguishable single leading entanglement spectral level when $D < L/4$, while this feature disappears in 
    general when $D \geq L/4$. 
    Note that the number of the nontrivial entanglement spectral levels in this case increases with $D$ as $4^D$.

\section{$D$ dependence of string order}  \label{app:str_order_D}

\begin{figure}[!ht]
  \centering
  \includegraphics[width=\linewidth]{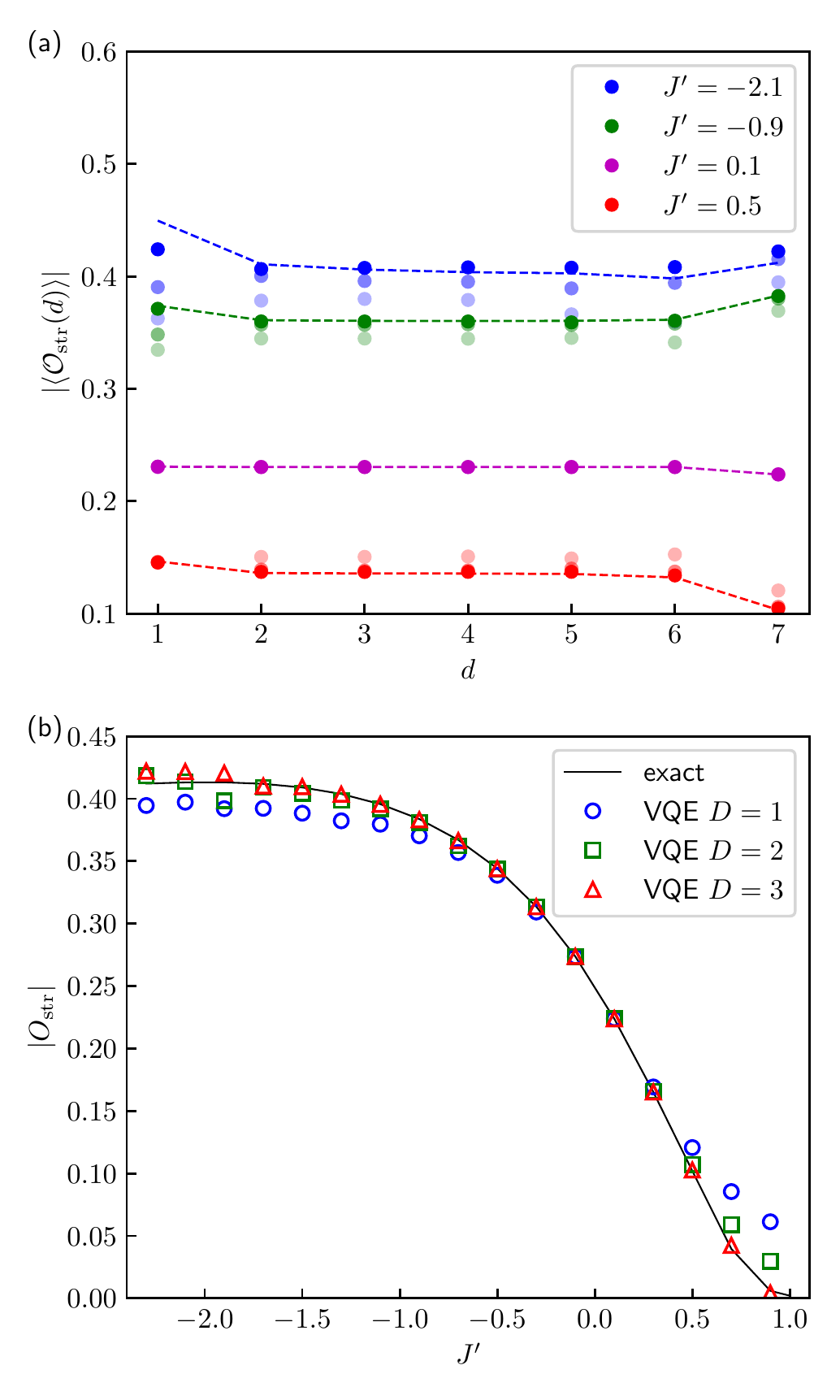}
  \caption{\label{fig:vqe_str_circ_depth_jps}
    Same as Fig.~\ref{fig:vqe_str_jps} but for the VQE optimized circuit \textit{Ans\"{a}tze} $|\Psi(\boldsymbol{\theta})\rangle$ 
    with $D = 1$, 2, and 3. 
    In (a), the results for different values of $D$ are denoted by different shaded color intensities proportional to $D$. 
    The results for $D=1$ are the same as those in Fig.~\ref{fig:vqe_str_jps}(a). 
    In (b), the system size is $L=12$. 
    }
\end{figure}

In Sec.~\ref{sec:haldane_phase}, we demonstrate that the shallow $D = 1$ \textit{Ansatz} can faithfully describe the SPT Haldane 
phase in a wide parameter region. Here, we examine numerically how the accuracy of the VQE calculation is further improved 
with increasing the circuit \textit{Ansatz} depth $D$. 
Figure~\ref{fig:vqe_str_circ_depth_jps}(a) shows the expectation value of string operator, 
$\langle \mathcal{O}_{\rm str}(d)\rangle$, evaluated numerically for 
the VQE optimized circuit \textit{Ans\"{a}tze} $|\Psi(\boldsymbol{\theta})\rangle$ 
with $D=1$, 2, and $3$. We can indeed confirm that, for a fixed $J'$, it approaches to the exact value with increasing $D$. 
Accordingly, as shown in Fig.~\ref{fig:vqe_str_circ_depth_jps}(b), 
the string order parameter $O_{\rm str}$ also becomes closer to the exact value, 
even for $J'$ away from $J'=0$ within $-2.5 < J' < 1$, 
indicating that more fine correlations of the ground 
state can be captured quantitatively by a deeper circuit \textit{Ansatz}.

\section{\label{app:vs_so4_brick_wall} Comparison with generic brick wall \textit{Ansatz}}

In this Appendix, we compare the results for the circuit \textit{Ansatz} $|\Psi(\boldsymbol{\theta})\rangle$ with 
those for a generalized brick-wall-type \textit{Ansatz} without considering the topological nature of the SPT Haldane phase. 
The latter \textit{Ansatz} constructs a variational state
\begin{equation}
  \label{eq:so4_brick_wall}
  |\Phi(\boldsymbol{\psi}, \boldsymbol{\theta}, \boldsymbol{\phi}, \boldsymbol{\alpha}, \boldsymbol{\beta}, \boldsymbol{\gamma})\rangle = \prod_{d = 1}^{D} \prod_{\{i,j\}_{d}} \hat{V}_{ij}(\psi_{ij}^{(d)}, \theta_{ij}^{(d)}, \phi_{ij}^{(d)}, \alpha_{ij}^{(d)}, \beta_{ij}^{(d)}, \gamma_{ij}^{(d)})|0\rangle~,
\end{equation}
where $\hat{V}_{ij}(\psi_{ij}^{(d)}, \theta_{ij}^{(d)}, \phi_{ij}^{(d)}, \alpha_{ij}^{(d)}, \beta_{ij}^{(d)}, \gamma_{ij}^{(d)}) \in \mathbf{SO}(4)$ 
represents a generic parametrized real rotation gate acting on qubits $i$ and $j$ at the $d$-th depth. 
Note that each $\mathbf{SO}(4)$ two-qubit gate $\hat{V}_{ij}$  
is characterized with six real parameters $\psi_{ij}^{(d)}$, $\theta_{ij}^{(d)}$, $\phi_{ij}^{(d)}$, $\alpha_{ij}^{(d)}$, $\beta_{ij}^{(d)}$, 
and $\gamma_{ij}^{(d)}$. 
The arrangement of these gates within the $d$-th depth is the same as that shown in Fig.~\ref{fig:eswap}(a). 
We denote this circuit \textit{Ansatz} as $\mathbf{SO}(4)$ brick wall \textit{Ansatz}. 
The number of parameters in $\mathbf{SO}(4)$ brick wall \textit{Ansatz} scales with $D$ as $6\times (L-1) \times D$. 
Note that since the ground state of the AHC can always be represented by a real wavefunction,  
utilizing a real state $|\Phi\rangle$ as an \textit{Ansatz}, instead of a circuit \textit{Ansatz} based on more general 
$\mathbf{SU}(4)$ two-qubit gates, is advantageous for reducing the variational search space.

\begin{figure}[t]
  \centering
  \includegraphics[width=\linewidth]{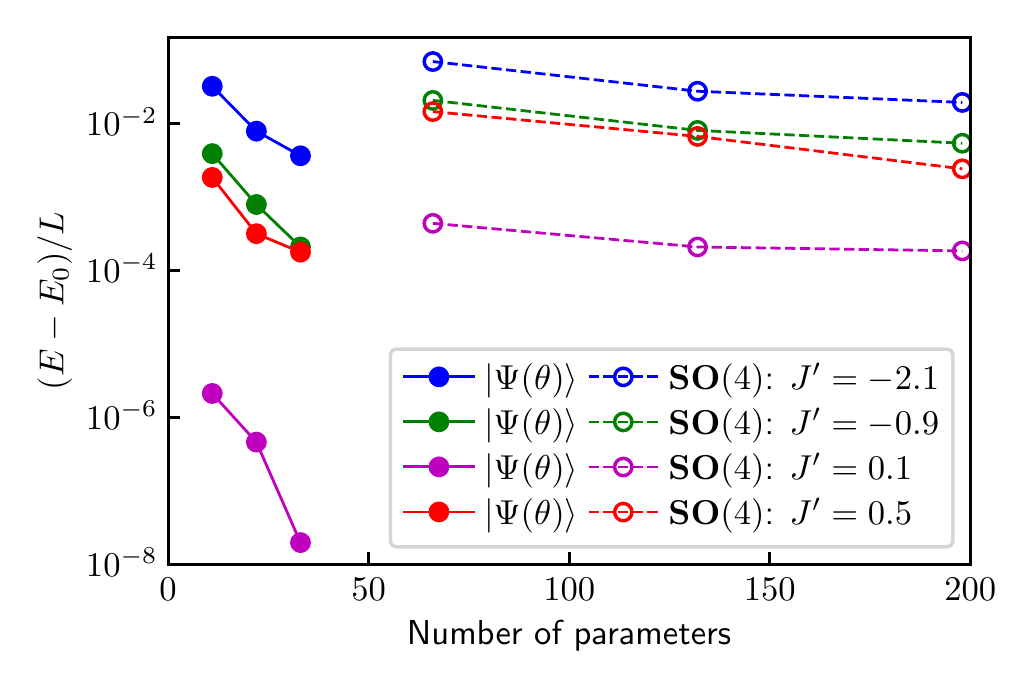}
  \caption{\label{fig:vqe_de0_eswap_vs_so4}
  The energy deviation of the VQE optimized circuit \textit{Ans\"{a}tze} $|\Psi(\boldsymbol{\theta})\rangle$ 
  and the VQE optimized $\mathbf{SO}(4)$ brick wall \textit{Ans\"{a}tze} 
  $ |\Phi(\boldsymbol{\psi}, \boldsymbol{\theta}, \boldsymbol{\phi}, \boldsymbol{\alpha}, \boldsymbol{\beta}, \boldsymbol{\gamma})\rangle$ 
  with $D=1$, 2, and 3 from the exact ground-state energy $E_0$ as a function of the number of variational parameters, which 
  increases with $D$ as $(L-1) D$ for $|\Psi(\boldsymbol{\theta})\rangle$ and $6 (L-1)  D$ for 
  $ |\Phi(\boldsymbol{\psi}, \boldsymbol{\theta}, \boldsymbol{\phi}, \boldsymbol{\alpha}, \boldsymbol{\beta}, \boldsymbol{\gamma})\rangle$. 
  These results are obtained for the $L=12$ AHC under OBC with four different values of $J'$ indicated in the figure.}
\end{figure}

We perform VQE simulations for typical values of $J'$ in the SPT Haldane phase using $\mathbf{SO}(4)$ brick wall \textit{Ans\"{a}tze} 
with $D = 1$, 2, and 3,  
and the results of the ground-state energy are compared with those for the VQE optimized circuit \textit{Ans\"{a}tze} 
$|\Psi(\boldsymbol{\theta})\rangle$ with $D = 1$, 2, and 3 in Fig.~\ref{fig:vqe_de0_eswap_vs_so4}. 
While the $\mathbf{SO}(4)$ brick wall \textit{Ansatz} has more variational parameters, the VQE optimized circuit \textit{Ansatz} 
$|\Psi(\boldsymbol{\theta})\rangle$ consistently obtains better energy with the deviation from the exact value at least one order 
of magnitude smaller than the $\mathbf{SO}(4)$ brick wall \textit{Ansatz}. 
Moreover, the convergence to the exact value with increasing D, i.e., the number of the variational parameters, is much faster 
for the circuit \textit{Ansatz} $|\Psi(\boldsymbol{\theta})\rangle$. 
This comparison clearly demonstrates the superiority of the circuit \textit{Ansatz} $|\Psi(\boldsymbol{\theta})\rangle$ proposed in this 
paper, which properly considers the nontrivial topology of the target state in the VQE for correlated 
topological phases.

\begin{thebibliography}{72}
\bibitem [{not()}]{noteparity}%
      \BibitemOpen
      \href@noop {} {}\bibinfo {note} {Note that $\hat{\mathcal{C}}_{0}$ in
      $|\phi_{s}\rangle$ has parity even and thus the bond central inversion
      symmetry is irrelevant in the sense that this symmetry cannot distinguish the
      SPT phase from a trivial state in this case.}\BibitemShut {Stop}%
\end{thebibliography}
\end{document}